\documentclass{emulateapj}

\def\arcdeg{\hbox{$^\circ$}}

\def\deg2{\hbox{$\rm deg^{2}$}}

\def\lsim{\mathrel{\rlap{\lower4pt\hbox{\hskip1pt$\sim$}}\raise1pt\hbox{$<$}}}                
\def\gsim{\mathrel{\rlap{\lower4pt\hbox{\hskip1pt$\sim$}}\raise1pt\hbox{$>$}}}                

\lefthead{Drake et~al.}
\righthead{Tidal Stream} 

\begin{document}
\title{Evidence for a Milky Way Tidal Stream Reaching Beyond 100 kpc} 

\author{
A.J.~Drake\altaffilmark{1}, M.~Catelan\altaffilmark{2,3}, S.G.~Djorgovski\altaffilmark{1}, 
G.~Torrealba\altaffilmark{2,3}, M.J.~Graham\altaffilmark{1},
A.~Mahabal\altaffilmark{1},\\
J.L.~Prieto\altaffilmark{5}, 
C.~Donalek\altaffilmark{1}, R.~Williams\altaffilmark{1}, S.~Larson\altaffilmark{6},
E.~Christensen\altaffilmark{6} and E.~Beshore\altaffilmark{6}
}

\altaffiltext{1}{California Institute of Technology, 1200 E. California Blvd, CA 91225, USA}
\altaffiltext{2}{Pontificia Universidad Cat\'olica de Chile, Departamento de Astronom\'ia y Astrof\'isica, 
Facultad de F\'{i}sica, Av. Vicu\~na Mackena 4860, 782-0436 Macul, Santiago, Chile}
\altaffiltext{3}{The Milky Way Millennium Nucleus, Av. Vicu\~{n}a Mackenna 4860, 782-0436 Macul, 
Santiago, Chile}
\altaffiltext{4}{Department of Astronomy, Princeton University, 4 Ivy Ln, Princeton, NJ 08544}
\altaffiltext{5}{The University of Arizona, Department of Planetary Sciences,  Lunar and Planetary 
Laboratory, 1629 E. University Blvd, Tucson AZ 85721, USA}

\begin{abstract}
  
  We present the analysis of 1,207 RR Lyrae found in photometry taken by the Catalina Survey's Mount Lemmon telescope.
  By combining accurate distances for these stars with measurements for $\sim$14,000 type-AB RR Lyrae from the Catalina Schmidt
  telescope, we reveal an extended association that reaches Galactocentric distances beyond 100 kpc and overlaps the Sagittarius
  streams system. This result confirms earlier evidence for the existence of an outer halo tidal stream resulting from a
  disrupted stellar system. By comparing the RR Lyrae source density with that expected based on halo models, we find 
  the detection has $\sim 8 \sigma$ significance.
  We investigate the distances, radial velocities, metallicities, and period-amplitude distribution of the RR Lyrae.
  We find that both radial velocities and distances are inconsistent with current models of the Sagittarius stream. We also
  find tentative evidence for a division in source metallicities for the most distant sources. Following prior analyses, we
  compare the locations and distances of the RR Lyrae with photometrically selected candidate horizontal branch stars and
  find supporting evidence that this structure spans at least $60\arcdeg$ of the sky. We investigate the prospects
  of an association between the stream and unusual globular cluster NGC~2419.

\end{abstract}


\section{Introduction}

The study of the formation, composition, mass and kinematics of galaxy halos are among the most active areas of modern
astrophysical research. The formation of galaxy halos are now widely believed to be due to hierarchical structure
formation (e.g., Freeman \& Bland-Hawthorn 2002) where galaxies are produced by the combination of a monolithic collapse
(Eggen et al.~1962) and the accretion of galactic fragments (Searle and Zinn 1978).  For the halo of the Milky Way, all
aspects of the halo can probed by studying the nature of the remnants of disrupted dwarf galaxies.

Numerous tidal streams and dwarf galaxies have been discovered within the Galactic halo in the last 20 years (e.g.,
Ibata et al. 1994; Ibata et al.~2001; Vivas et al.~2001; Grillmair 2006; Newberg et al.~2009).  The most well-studied of
these is the Sagittarius dwarf galaxy (Sgr, Ibata et al. 1994) and its associated tidal stream. The Sgr streams have
been traced on large scales using blue horizontal branch (BHB) stars (Newberg et al. 2003), M-giants (Majewski et al.
2003), and main-sequence turn off (MSTO) stars (e.g., Belokurov et al. 2006; Koposov et al.~2012).  The Sgr system has
also been studied using RR Lyrae (RRL, Vivas \& Zinn 2006; Miceli et al.~2008; Sesar et al.~2010).  Other halo streams
and structures discovered recently include a Virgo stellar stream (VSS; Vivas \& Zinn 2006; Vivas et al. 2008), a Virgo
overdensity (VOD, Newberg et al. 2002; Newberg et al.~2007), an overdensity in Pisces (Sesar et al.~2007; Kollmeier et
al.~2009) and a Monoceros stream (Newberg et al. 2002; Majewski et al.~2003).  Additionally, a Cetus stream has been
discovered in the south (Newberg et al. 2009; Koposov et al.~2012) and also evidence for an overdensity dubbed 
the Hercules-Aquila Cloud (Belokurov et al.~2007).

Although the Sgr stream system has been the focus of much study, since the structure is extends completely around the
Galaxy, it is yet to be fully mapped in velocity, metallicity and distance. Furthermore, the cause of the bifurcation in
Sgr stream stars discovered by Belokurov et al.~(2006) is yet to be explained. Additionally, Newberg et al.~(2003)
proposed the existence of a stream of stars associated with the Sgr dwarf galaxy at Galactocentric distances of $\rm
90\, kpc$.  These authors photometrically identified candidate BHB stars in SDSS data and found evidence for an
overdensity of stars with $g \sim 20.5$.  This overdensity was found using SDSS data covering $110\arcdeg < \alpha < 130
\arcdeg$, $20\arcdeg < \delta < 50\arcdeg$.  Newberg et al.~(2003) attributed this feature to a trailing arm of the Sgr
stream, yet also noted a possible link to the globular cluster NGC~2419. Their analysis showed evidence for a feature
visible across $\sim 20 \arcdeg$ along the Sgr plane.  Newberg et al.~(2007) continued this work and found additional
evidence for what they called the Sgr trailing tail using BHB candidates selected from SDSS Data Release 5 (DR5;
Alderman-McCarthy et al.  2007).  More recently, Ruhland et al.~(2011) found additional evidence for this overdensity
using candidate BHB stars selected from the larger SDSS DR7 dataset. As with Newberg et al.~(2003, 2007), these authors
found a stream of BHB candidates to be located at heliocentric distances of 60-80 kpc with $3.8 \sigma$ significance.
Within their analysis the feature was seen to span $\sim 90\arcdeg$ of the sky in the region $110 < \alpha <
200\arcdeg\!$. Ruhland et al.~(2011) also compared this extended stream with five sets of published numerical
simulations of the Sgr stream. They found that none were able to explain the existence of Sgr stream stars in the
location observed.

Current evidence for an outer Sgr stream/tidal-tail remains based purely on photometrically selected BHB stars. As
demonstrated by Sirko et al.~(2004) and Ruhland et al.~(2011), it is not possible to select a 100\% pure set of BHB
stars using SDSS photometry alone. Depending on the colour cuts used one has to accept varying levels of contamination
blue straggler (BS) stars which far out number BHB stars at the greatest detection completeness. 
With strict colour cuts, that remove a large fraction of the BHB stars, the level of BS star contamination can be reduced 
to 30\% (Ruhland et al.~2011).  However, to retain a large number of BHB stars, a 50\% or greater contamination level must
be accepted (Brown et al.~2005; Bell et al.  2010).

In contrast to photometric selection, it is possible to select BHB stars with much less contamination by combining
colour selection with high S/N spectra because BHB stars have significantly lower surface gravity than the BS
stars (Yanny et al.~2000).  The SDSS SEGUE-I and SEGUE-II projects (Yanny et al.~2009) undertook spectroscopy of $\sim$
300,000 star with an aim to identifying BHB stars and trace Galactic halo structure. However, while radial velocities
are available for stars to $g \sim 20.3$ (Yanny et al.~2009), the S/N required to accurately measure $\log({\tt g})$ 
limits the survey to BHB stars with $g < 19.5$ ($d\sim60$ kpc ). Almost all of the BHB candidates discovered by Newberg et al.~(2003,
2007) and Ruhland et al.~(2011) are beyond this limit and do not have measurements of $\log({\tt g})$ to 
confirm their nature.  Additionally, while the SDSS SEGUE surveys cover $4300\,deg^2$, the fields cover a patchwork over 
the Northern sky, rather than just the Sgr stream region where this feature is located.

Like BHB stars, RRL stars also exist on horizontal branch (HB). However, unlike BHB stars,
RRL exhibit a significant level of intrinsic photometric variability. Based on their characteristic variability
RRL can be cleanly separated from other stars. Type-ab RRL stars (RRab) have absolute magnitudes of $M_V=0.6$ with
uncertainties of 6\% (Catelan et al.~2009). This makes them excellent distance indicators.
Although, significant numbers of repeated observations are required 
to provide accurate average magnitudes.

To date, a few tens of thousands of RRL are known in dense regions near the Galactic bulge and in the Magellanic clouds
due to microlensing surveys (Soszy\'nski et al. 2009, Pietrukowicz et al. 2012). Recently we discovered $\sim$ 10,000
RRab over a large fraction of the sky (Drake et al. 2013; DR13).  However, even with the DR13 RRab's most of the
Galactic halo remains unexplored at heliocentric distances $\rm > 60 kpc$.

In this paper we outline our search, discovery, and calibration of RRL at distances up to 100 kpc and beyond.  We will
compare our discoveries with results from past surveys and undertake an analysis of the distant HB stars uncovered.

\section{Observational Data}

The Catalina Sky Survey began in 2004 and uses three telescopes to cover the sky between Declination $-75\arcdeg$ and
$+65\arcdeg$ in order to discover Near Earth Objects (NEOs) and Potential Hazardous Asteroids (Larson et al.~2003).  The
three telescopes are each considered sub-surveys.  These consist of the Catalina Schmidt Survey (CSS), the Mount Lemmon
Survey (MLS) and the Siding Spring Survey (SSS).  In addition to avoiding high declinations, the Galactic plane region
is avoided by between 10 and 15 degrees due to reduced source recovery in crowded stellar regions.  All of the survey
images are taken in sequences of four observations separated by ten minutes, and all the observations are unfiltered to
maximize throughput. Photometry is carried out using the aperture photometry program SExtractor (Bertin \& Arnouts
1996).  All the Catalina data is also analyzed for transient sources by the Catalina Real-time Transient Survey
(CRTS\footnote{http://crts.caltech.edu/}, Drake et al.~2009; Djorgovski et al.~2011).

In paper-I (DR13), we concentrated on analyzing RRab data from the CSS sub-survey.  In this paper we mainly work with
data observed by the MLS 1.5m telescope. This data predominately covers Ecliptic latitudes $-10\arcdeg < \beta < 10
\arcdeg$.  For this telescope each image from the $4{\rm k} \times 4{\rm k}$ Catalina CCD camera covers 1.1 $\rm deg^2$
on the sky.  Observations are taken 21 nights per lunation avoiding bright time. Typical exposures of 30 seconds to
reach $V \sim 21.5$. In total, the MLS source catalog consists of 155 million sources.

\subsection{Calibration}

As noted in DR13, the MLS data has the same photometric sensitivity as the CSS data since all observations are taken
unfiltered with the same type of CCD camera. All images are processed with the same software. From DR13, the colour
transformation from Catalina photometric system $\rm V_{CSS}$ magnitudes to Johnson $V$ is given by:

\begin{equation}\label{tran}
V = V_{CSS} + 0.31 \times (B-V)^2 + 0.04.\\
\end{equation}

The average B-V colour of RRL is about 0.3 mag with stars varying between about 0.1 and 0.5 as they pulsate (e.g.,
Clube et al. 1969; Stepien 1972; Cacciari et al. 1987; Layden 1997; Nemec 2004). As the MLS data has no colour
information, we adopt the average values in our analysis.  This gives rise to an maximum uncertainty of $\sim 0.07$ mags
in the $V$ magnitudes.  Combining this with the photometric uncertainty, we expect a colour-based dispersion of $\sigma
= 0.09$ mags in our RRL photometry.

\subsection{RR Lyrae selection}

To find RRL candidates we follow the analysis undertaken in DR13. That is, we first selected variable sources using a
Welch-Stetson variability index $I_{\rm WS}$ (Welch \& Stetson 1993).  3.1 million MLS sources were selected as variable
candidates using $I_{\rm WS} > 0.6$.  All sources were processed for periodicity using the Lomb-Scargle (LS, Lomb 1976;
Scargle 1982) periodograms.  In total $170$ thousand sources were found to exhibit significant periodicity with a false
alarm probability $p_0 < 1\times 10^{-5}$.  However, the bulk of these detections are due to the systematic sampling
of the data and occur at integer day frequencies.

Here, as with DR13, we are primarily interested in RRab's since they have well known magnitudes and characteristic light
curves that are not easily mistaken for other periodic variables.  After excluding the period aliases near 0.5 and 1 day,
3087 of the objects found to have periods between 0.34 and 1.4 days. This period region was
deliberately chosen to be broader than the RRab range in order to include sources found by the LS technique at aliases
of their true periods.

As with the CSS sources of DR13, each of the RRab candidate was run through the AFD software (Torreabla et al.~2012)
to select the best period from among the best 15 given by the LS and AoV software (Schwarzenberg-Czerny 1989).  Briefly,
this process involves Fourier fitting and iteratively rejecting bad data to determine the best period based on reduced
$\chi^2$ values.  Non-RRab sources are rejected using the M-test statistic (Kinemuchi et al,~2006; Equation 8), as well 
as the Fourier fit order. This initial selection resulted in the detection of 1125 RRab candidates.

\subsection{Faint RRab stars}

\begin{figure}{
\epsscale{1.0}
\plotone{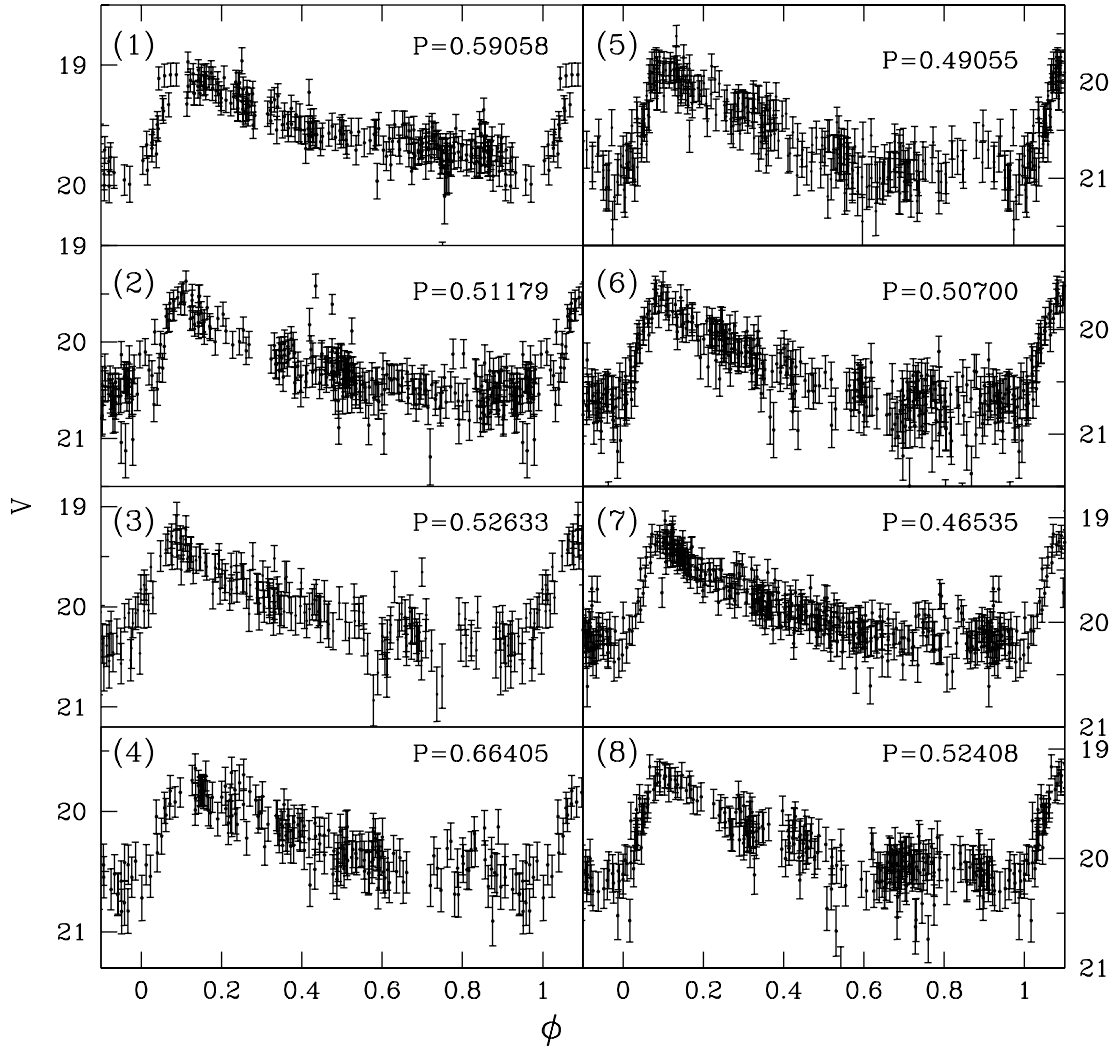}
\caption{\label{MLSRRab}
The period-folded light curves of eight of the 
Galactic halo RRab discovered in MLS photometry. Here the numbered 
figures are correspondingly:
(1) MLS J093311.1+131726; (2) MLS J092028.4+153244;
(3) MLS J050243.0+203816; (4) MLS J073511.7+185253; 
(5) MLS J080834.1+192509; (6) MLS J081338.1+191449;
(7) MLS J074044.7+204658 \& (8) MLS J043725.2+211020.
}
}
\end{figure}

Upon reviewing the light curves of the RRab candidates we discovered a number of faint RRab in the MLS data with $V\sim
20.5$. Such RRL were found to be mainly concentrated in the Gemini constellation. In Figure \ref{MLSRRab}, we plot the
phased light curves of eight of these faint stars.  These distant sources are very important for defining distant streams
within the Galactic halo. From simulations, based on adding artificial RRab light curves to our data, we discovered that
our detection sensitivity was $< 50\%$ for RRab's with $V > 20$.

To improve our completeness we investigated the light curves of MLS sources with SDSS spectroscopic types A0 and F5 that
had $19 < V < 21$. In addition, we used SDSS photometry to select stars with the same colour range as the RRab
stars that we had discovered.  These stars were matched with MLS sources at lower variability and periodicity 
thresholds ($I_{\rm WS} > 0.3$ and $p_0 < 0.01$) to select RRab candidates.  To better constrain the variability, 
we combined the MLS photometry with shallower CSS data.  Likewise, we searched the CSS data for additional faint 
RRab candidates by selecting objects with RRab colours in SDSS photometry. 
Furthermore, since our analysis in DR13 showed that $\sim 17\%$ of RRab's were missed because the 
selection based on LS periods, we reprocessed the CSS and MLS photometry using the AoV period-finding software.

We inspected the light curve of each RRL candidate to assess variability and then carried out period finding for all the new
variable candidates.  In total, approximately six thousand additional sources were searched.  Each of the new sources
exhibiting some sign of periodicity were searched for improved periods using our Fourier-fitting AFD 
process. In a number of cases the RRL were found to exhibit Blazhko phase and amplitude variations (Blazhko 1907).

In total the additional searches yielded an 219 extra MLS RRL and 2051 CSS RRab's. Of these only 425 were previously
known.  As distant RRL are important for defining Halo streams we retain the 17 clear c-type RRL (RRc) that were
discovered in MLS data beyond 70 kpc. A number of these have SDSS spectra.  After removing non-RRL sources the final 
set of MLS RRL consists of 1,207 stars and is presented in Table 1. The new RRab's found in CSS data are given in Table 2.

\begin{figure}{
\epsscale{0.7}
\plotone{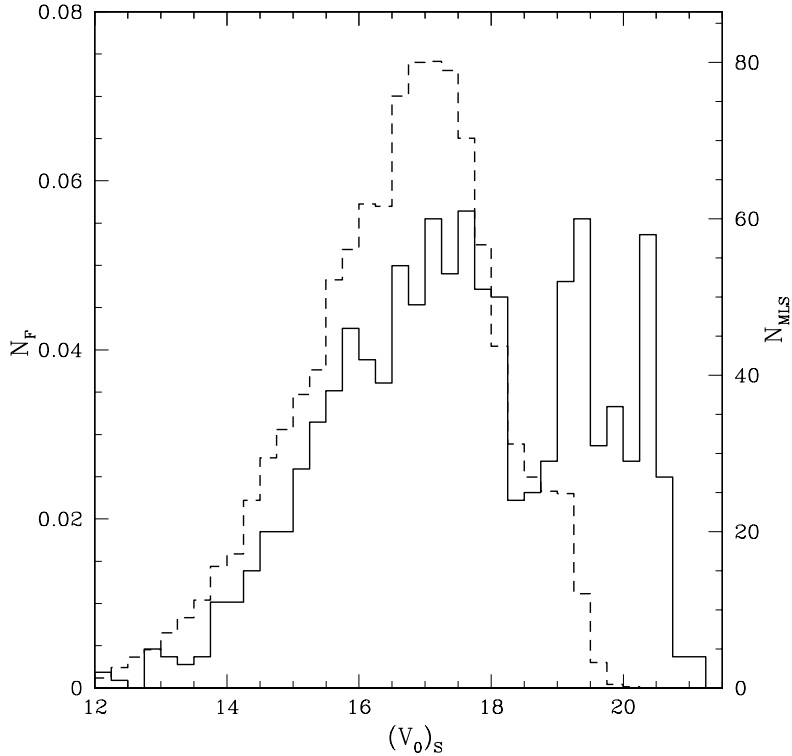}
\caption{\label{MLSHist}
  The distribution of RRL magnitudes.  The fractional number ($N_F$) of RRL static magnitudes for the 1,207 MLS RRLs
  found in the survey are plotted with the solid line. The $\sim 14500$ CSS RRab's are plotted with the dashed line. The
  peak in the MLS distribution near $V=19.5$ is mainly due to RRL in the two separate arms of the Sgr stream system. The
  peak near mag $V=20.5$ is mainly due to the distant Gemini feature. The actual numbers of MLS RRL are noted by
  $N_{MLS}$.  The actual number of CSS RRab's are $N_{CSS} \sim 13.4 \times N_{MLS}$.
}
}
\end{figure}

In Figure \ref{MLSHist}, we present the magnitude distribution of MLS and the CSS RRL.  The MLS histogram shows clear bumps
near $V=19.5$ and $V=20.5$.  The first bump is easy to understand as it is clearly due to the Sagittarius leading and
trailing arms (as seen in DR13).  The second peak is due to the more distant RRL that are concentrated near the Galactic
anti-center. We hereafter call this the Gemini stream. In section \ref{SagStream}, we will investigate the origin of
this feature.

\section{RRab Distances}

The absolute magnitudes of RRab's are given by Catelan \& Cort\'es (2008):

\begin{equation}
M_V = 0.23 \times {\rm [Fe/H]}_{\rm ZW84}+ 0.948,
\end{equation}

\noindent where ${\rm [Fe/H]}_{\rm ZW84}$ is the metallicity in the Zinn \& West (1984) scale.
The average metallicity for RRab's with SDSS spectra found by DR13 was $\rm [Fe/H]= -1.55$.  The comparison given in
DR13, between distant halo RRab's ($d_G > 33.5$ kpc) and the full set of RRab's showed little difference in the
distribution or average metalicity.  However, CSS RRab's with SDSS metallicities are stars with $V > 16$,
corresponding to distance of $>12$ kpc. Brighter foreground field RRab's are likely to have higher
metallicities.  The MLS RRab's should have a similar metallicity distribution to the larger sample of CSS sources
discovered in DR13.  Therefore, we adopt the average CSS metallicity and assume RRab's have magnitude $M_V=0.59$.
This value is close to the value of 0.6 that is often adopted (e.g., Keller et al.~2008, Sesar et al.~2010).  The RRL V
magnitudes were corrected for extinction using Schlegel et al.~(1998) reddening maps.  The dispersion in the metallicity
from DR13 is 0.3 dex, which corresponds to a variation of 0.07 magnitudes. The distances to individual sources are
determined using:

\begin{equation}
d = 10^{((V_0)_{\rm s} - M_V + 5)/5} 
\end{equation}

\begin{figure}{
\epsscale{1.0}
\plotone{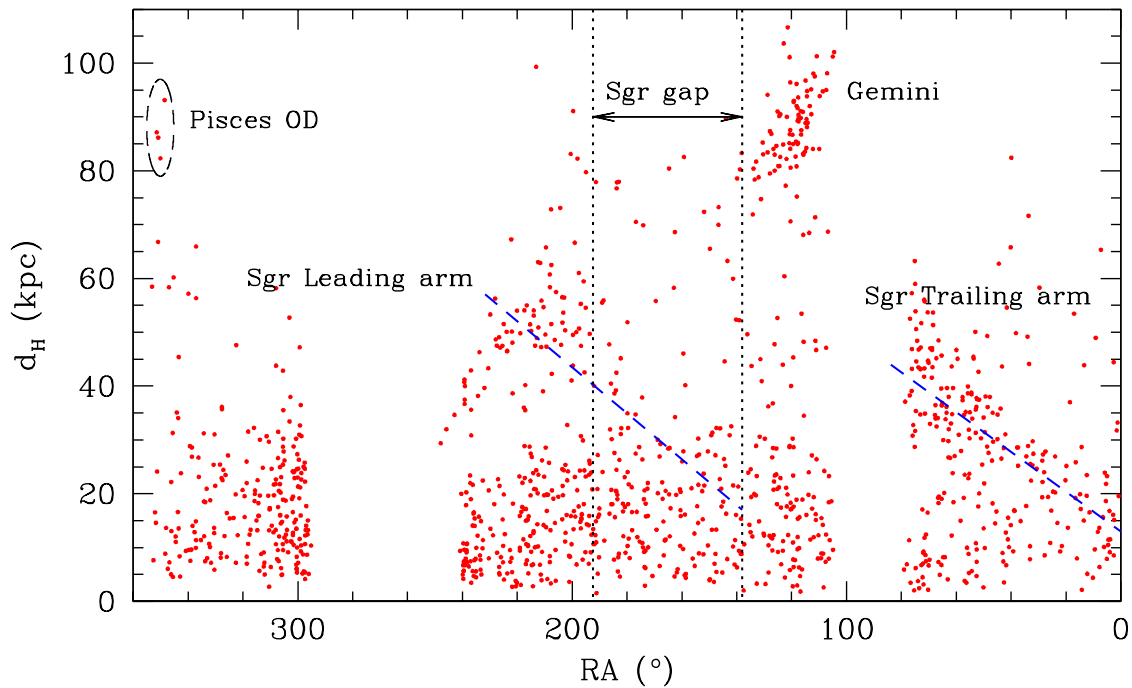}
\caption{\label{Dist}
  The heliocentric distance distribution for all the MLS RRab stars presented in this work. Each of the distances is
  derived from $V$ magnitudes corrected to static star values.  Known features including the Sagittarius streams and
  Pisces overdensity (Pisces OD; Sesar et al.~2007) are marked.  We also also marked the central region of the Sgr
  streams that is not covered by the MLS data (Sgr gap).
}
}
\end{figure}

Here we correct the average RRab $V_0$ magnitudes to static values $(V_0)_{\rm s}$ using values derived from a
polynomial fit to the amplitude corrections given by Bono et al.~(1995).  Combining the uncertainties from the
photometric calibration and colour variation of 0.09, with the variation in metallicity we derive an overall uncertainty
of 0.11 magnitudes, corresponding to a $\sim 5\%$ uncertainty in distance. As demonstrated by DR13, faint RRL have
larger uncertainties in their average magnitude. However, these uncertainties should generally not exceed 0.2 magnitudes
($\sim 10\% $ in distance).  The faintest MLS RRab in our dataset have $(V_0)_{\rm s} \sim 21$ corresponding to 
heliocentric distances of 120 kpc. In Figure \ref{Dist}, we plot the distribution of the distances of MLS RRL
and mark the locations of known halo features as well as the Gemini stream.

\section{Comparison with known RRab}

\begin{figure}{
\epsscale{0.7}
\plotone{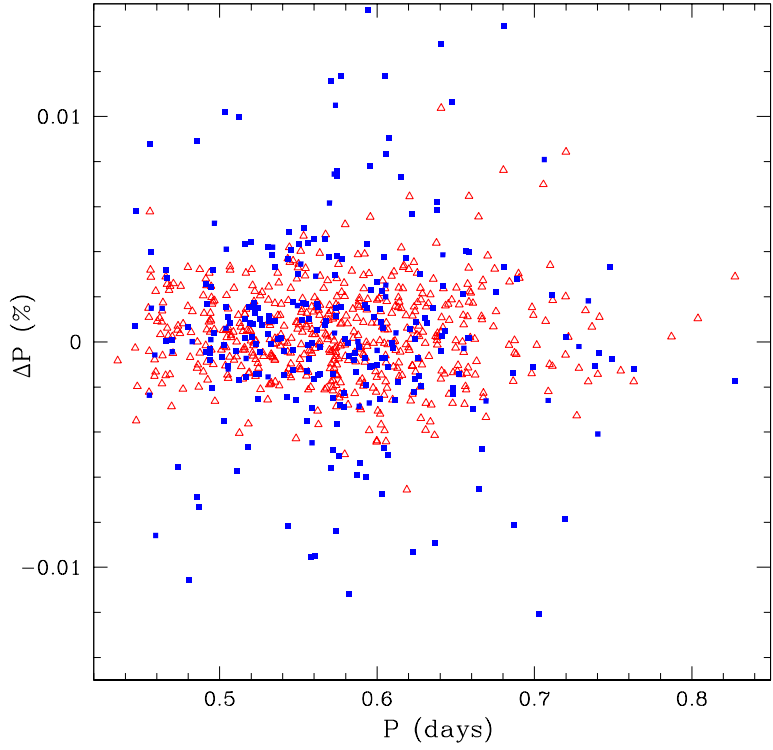}
\caption{\label{Other}
  Comparison of MLS RRab periods.  Here we plot the difference in period between that derived from MLS data and
  elsewhere for previously known RRab's. Periods based on the Drake et al.~(2012) CSS RRab's catalog are given with
  triangles, while all other sources are presented as filled boxes.
}
}
\end{figure}

We matched the MLS RRab's with past RRL discoveries using the SIMBAD and the International Variable Star Index (VSX;
Watson et al.~2006) databases, as well as those presented by DR13. In total, there were 605 matches with past surveys of
which 316 came from CSS data (DR13) and 289 were discovered in other surveys. As there is overlap between CSS and the
other previously known RRab's, there was CSS photometry for 540 of the MLS RRab's. Of the 289 SIMBAD and VSX RRab's, 278 of the
RRab have known periods given by VSX. In Figure \ref{Other}, we compare the periods derived from MLS data with previous
determinations.  The figure shows that the agreement between the CSS and MLS periods is much better than with other
sources.  Assuming the differences in periods are normally distributed, the MLS-CSS matches have $\sigma= 0.0020\%$,
while the MLS-VSX source matches have $\sigma=0.0042\%$.  As the level of agreement is good in both cases, we have
confidence in the MLS RRab periods.

\subsection{Comparison with SEKBO RRL Candidates}

The SEKBO survey discovered 2016 candidate RR Lyrae in a survey covering 1675 $\rm deg^2$ along the ecliptic (Keller et
al.~2006). Like the MLS survey, the SEKBO survey covered a band within $10\arcdeg$ of the ecliptic.  However, SEKBO RRL
candidate selection was based on the two colours observed simultaneously by the MACHO camera and variability was
determined from between 2 and 8 epochs of images.  In contrast, the MLS RRab candidates are observed unfiltered and have
an average of 130 observations (sampling ranges from 42 to 342 measurements).

We matched the MLS RRab with the SEKBO RRL candidates and found only 103 matches.  However, this is not unexpected since
the sampling of MLS is highly concentrated within few degrees of the ecliptic with few observations having been taken near
the $\sim 10\arcdeg$ limit. For example, 70\% of the MLS RRab's are within $2\arcdeg$ of the ecliptic, compared 
to 27\% of the SEKBO RRL candidates.

To extend the comparison, we also matched the SEKBO candidates with RRab's from the DR13 sample and the SSS RRab's
(Torrealba et al.~2012). This yielded an additional 538 RRab matches. Because of overlap between the CSS, MLS and SSS
RRab catalogs there were 540 unique RRab stars matching SEKBO RRL candidates.  Clearly this number falls far short of
the number of SEKBO candidates.

\begin{figure}{
\epsscale{0.7}
\plotone{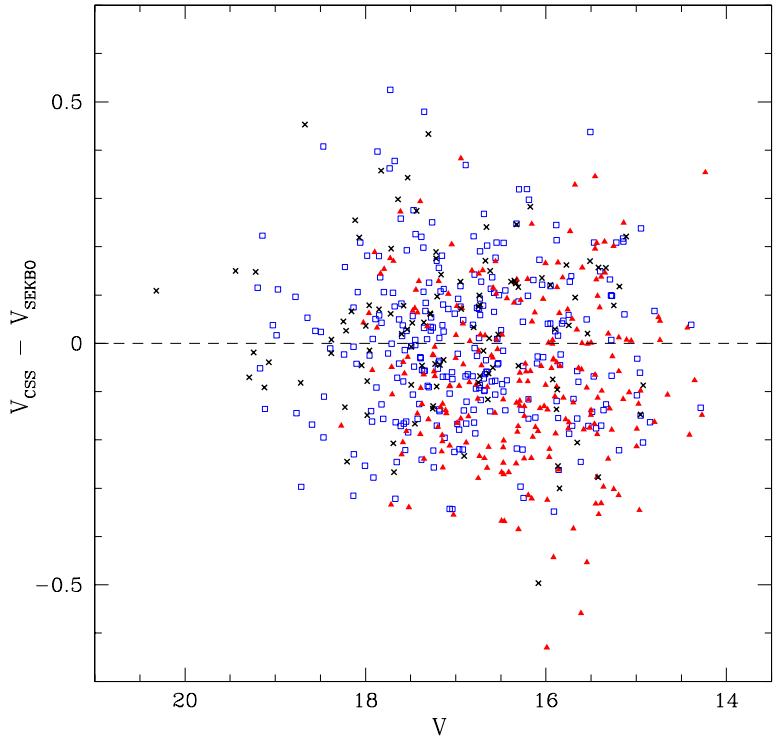}
\caption{\label{sekbo}
  The difference between V magnitudes from CSS and from SEKBO. The black crosses are the MLS matches, the blue boxes are
  CSS matches and red triangles are SSS matches.
}
}
\end{figure}

We compared average V magnitudes for the Catalina sources with those of the SEKBO sources.  In Figure \ref{sekbo}, we
present the V magnitude differences between the measurements.  The average difference between magnitudes is -0.02
magnitudes with $\sigma_{V} = 0.16$ magnitudes, corresponding to a $7.6\%$ uncertainty in source distances. The internal
dispersion for CSS RRab magnitudes from overlapping fields is $\sigma =0.04$ mags and thus does not significantly
contribute to the observed dispersion.

In order to investigate the SEKBO candidates not found in the Catalina RRab catalogs, we matched the remaining 1481
unmatched SEKBO objects to the SDSS DR8 spectroscopic catalog. We found 40 matches, of these 35 had measured $\log({\tt g})$
values and 23 had $\log({\tt g}) < 3.75$ consistent with RRL (DR13). We also matched the SEKBO sources with SDSS DR8
photometry. Of the 276 sources with SDSS photometry, approximately one third had SDSS colours beyond the limits
observed for DR13 RRab's.  Next, we matched the 1481 SEKBO sources to photometry from MLS, CSS and SSS surveys. 
Among the matches, we sub-selected the objects exhibiting significant variability based on the $I_{\rm WS} > 0.6$ and also 
objects with LS periodic significance $p_0 < 1\times 10^{-5}$.
In total, we find matches for 90\% of the SEKBO objects, including 88\% of the sources that were not selected 
as RRab's. Combining the known Catalina RRab's and other sources with apparent variability, we find that 74.5\% of
the 1833 SEKBO sources we cover exhibit significant variability. In Table 3, we present the numbers of
matches with each of the Catalina surveys, as well as the number of objects selected by our variability
and periodicity significance thresholds.

For each of the 1314 SEKBO RRL candidates with Catalina photometry, that were not in our RRab catalogs, we determined a
LS period (regardless of their $I_{\rm WS}$ variability).  We inspected both the observed and phase folded light curves
of each source and determined a classification based on this photometry. As there are 2759 Catalina light curves matching
the 1314 SEKBO candidates, the majority of these sources have multiple light curves.  We discovered 551 additional RRL
candidates, 140 eclipsing binary candidates (mainly WUMa) and 140 other types of variable sources (including QSO, CVs,
$\delta$ Scuti variables and objects where the classification was unclear).  This discovery of many non-periodic sources
is not unexpected since a number of these are already known, including QSOs and cataclysmic variables.  For example,
SEKBO 106646.2532 (CSS080623:140454-102702) is a CV that was discovered in outburst by CRTS in 2008 (Kato et al.~2012).
Additionally, based on SDSS DR8 spectroscopy, SEKBO-105832.627 is a CV, SEKBO-117146.1412 is a QSO at z=1.02, and
SEKBO-096514.1360 is an eclipsing WD-MD binary system (Rebassa-Mansergas et al.~2010).

In order to obtain the best periods for the clear RRL, and to find RRL that may have been missed during our inspection,
we ran every light curve through the AFD software (Torrealba et al.~2012). For 307 of the 551 RRL candidates we found
good Fourier fits and periods consistent with RRab's and RRc's.  In Table 4, we present all the new SEKBO RRL found 
using Catalina data (including RRL candidates where the period is uncertain). This table contains 263 RRab, 282 RRc, 
and 11 RRd candidates.  

Combining the new RRab's with the 540 from Catalina catalogs we find that $\sim 36\%$ were missing from our catalogs in
good agreement with the simulations presented in DR13.  The overall sample suggests that 24\% of the RRL are RRc stars
whereas Keller et al.~(2006) estimated that $\sim 10\%$ of the SEKBO candidates would be RRc's. However, we note that
Pietrukowicz et al.~(2012) found $30\%$ RRc among their $\sim 15000$ bulge RRL. Nevertheless, it is possible that some
fraction of the RRc candidates we identify may be W~UMa variables (due to the similarity of their light curves).  The
uncertain separation of RRc's and W~UMa types is the main reason why RRc's are often excluded as distance indicators.

Our results suggest that $60\%$ of the SEKBO RRL candidates are likely RRL based on the 1833 objects covered by Catalina
data.  This is $2.2\sigma$ smaller than the $24\pm7\%$ non-RRL contamination estimated by Prior et al.~(2009a).
However, we note that the Prior et al.~(2009a) sample was based on an average of just nine photometric measurements for
$106$ ($\sim 5\%$) of the SEKBO stars.  Prior et al.~(2009b) followed 21 additional SEKBO RRLs overlapping the Sgr tidal
stream. More recently, Akhter et al.~(2012) followed 137 SEKBO candidates and found 57 to be RRLs, although they found a
high fraction with colours matching the colours of RRL candidates from SDSS data.

From Catalina photometry we have an average of 121 observations per light curve (250 per object) covering 90\% of the SEKBO
sources. Our analysis suggests that $\sim 25\%$ of the SEKBO sources are either not variable or have significantly
less variability than expected for RRL. Nevertheless, with $>1000$ likely RRL from the SEKBO survey, the results shows
that large numbers of RRL can be found using very small numbers of observations and colour selection.

\section{Outer Halo RR Lyrae}\label{SagStream}

The outer halo RRL discovered in this analysis overlap with tidal stream first identified by Newberg et al.~(2003)
using photometrically selected BHB candidates. As noted above, additional BHB candidates have since been found 
is SDSS DR5 (Newberg et al.~2007) and SDSS DR7 data (Ruhland et al.~2011). Furthermore, Ivezic et al.~(2004) discovered 
many RRL candidates overlapping the Sgr stream plane. The most distant sources coincide with the Newberg et al.~(2003) 
BHB candidates. However, the Ivezic et al.~(2004) RRL sample was based on photometric selection combined with 
limited variability information. Of the 1,269 Ivezic et al.~(2004) RR Lyrae candidates selected in the SDSS stripe-82 
region, only 483 (38\%) were eventually confirmed by Sesar et al. (2010). It seems likely the a similar fraction of 
the sources found in this marginally significant ($2-3\sigma$) detection by Ivezic et al.~(2004) were indeed RRL.

Recently, additional evidence for a distant group of RRL stars was found by Sesar et al.~(2012a). These authors
discovered eight RRab between $124\arcdeg < \alpha < 133\arcdeg$ and $18\arcdeg < \delta < 24\arcdeg$ at Heliocentric
distances $77 < d_{H} < 96$ kpc.  This region lies on the edge of the fields where BHB candidates were earlier
discovered by Newberg et al.~(2003).  Sesar et al.~(2012a) selected potential RRab candidates based on variable Palomar
Transient Factory (PTF) sources with RRL colours based on SDSS DR8 photometry.  Using simulations Sesar et al.~(2012a)
estimated that in their worst-case scenario their search recovered 95\% of their simulated sources, with at most one
RRab star being missed.  Six of the Sesar et al.~(2012a) RRab's were independently discovered in our searches, while the 
remaining two were missed due to poor sampling. For all six known RRab, we find periods matching those given by 
Sesar et al.~(2012a) to better than $0.04\%$.

\begin{figure}{
\epsscale{1.0}
\plotone{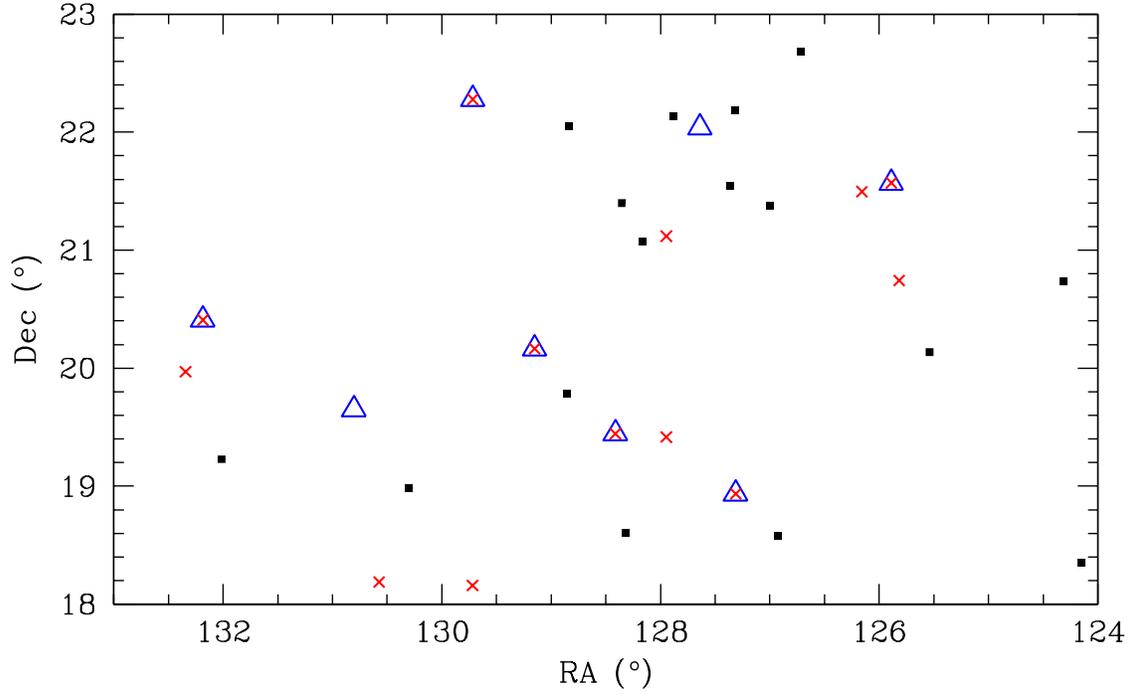}
\caption{\label{PTF}
  The distribution of RRab within the PTF Praesepe region.  The triangles show the locations of Sesar et al.~(2012a)
  RRab's, while the crosses mark RRab's found by MLS with $76 < d_H < 97$ kpc.  The dots mark the locations of RRab's
  with $d_H < 76$ kpc.
}
}
\end{figure}

In addition to the six PTF RRab, we have discovered seven RRab within the region and distant range noted by Sesar et al.~(2012).
We find 16 more RRab's in this region at heliocentric distances $d_H < 70$ kpc. As we missed two of the PTF RRab's it is
possible that there are even more RRab's within this region. In Figure \ref{PTF}, we present the locations of the RRab's
found within the PTF Praesepe fields.  Among the set of eight RRab's, Sesar et al.~(2012) found tentative evidence for
two separate groups of RRL based on radial velocities. One group of four RRab's was noted as having radial velocities in
the Galactic standard of rest of $\rm \langle Vgsr \rangle = 16 \pm 7 km s^{-1}$, while the other had $\rm \langle Vgsr
\rangle = 78 \pm 6\, km s^{-1}$. Although Sesar et al.~(2012) note that there is a 37\% chance that these stars are
drawn from the same Gaussian velocity distribution based on the Shapiro \& Wilk (1965) SW statistic, they assume that
their eight sources are representative of the overall velocity distribution and via numerical simulations find a $<
0.6\%$ chance that the sources would exhibit the observed velocities. These authors also discovered that the
metallicities of the two groups were consistent to within uncertainties. 

In total, we identify 103 RRab's at heliocentric distances beyond 70 kpc. Four of these RRab's appear near $\alpha =
355\arcdeg$, $d_H = 85$ kpc. These stars are associated with the Pisces stream (Sesar et al.~2007; Watkins et al.~2009).
However, we see no evidence for variations in distance. In agreement with the results of Kollmeier et al.~(2009). This
supports evidence that this system is not part of a tidal stream itself. Of the remaining stars, 82 RRab's lie within
the range $100\arcdeg < \alpha < 150\arcdeg$ and $14\arcdeg < \delta < 30\arcdeg$. These objects exhibit a range of
distances and thus comprise significant evidence for a stream of stars, suggesting association with a tidally disrupted
galaxy.

\begin{figure}{
\epsscale{1.0}
\plotone{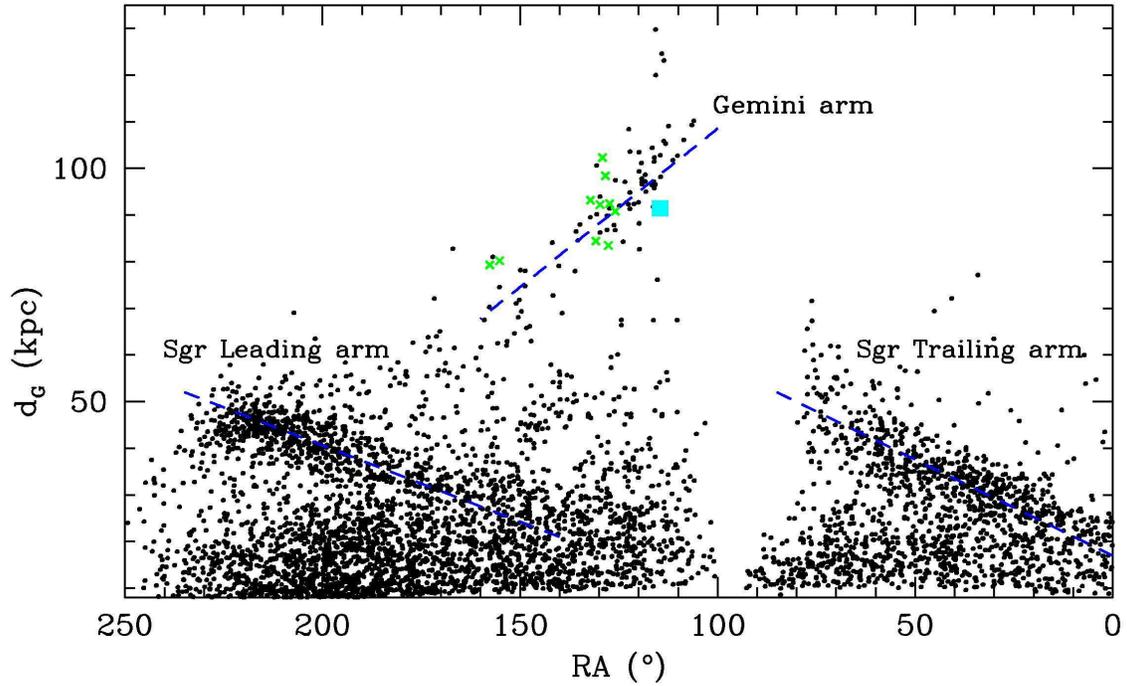}
\caption{\label{DistG}
  The Galactocentric distance distribution for stars within $11\arcdeg$ of the Sagittarius plane. Large red dots give the
  MLS data and black points give CSS RRab data. Triangles give newly discovered CSS RRab's and crosses give PTF RRab's. The dashed
  line presents a simple linear fit to the RRab with $d_G > 70$ kpc. The filled-box presents the distance and location
  of the globular cluster NGC 2419.  
} 
}
\end{figure}

To better understand the nature of the relation between the distant RRab's and the Sgr stream we selected the RRab
sources within $11 \arcdeg$ of the Sgr plane (as defined by Majewski et al.~2003).  Based on Koposov et al.~(2012) and
DR13. Most Sgr stream stars, including RRab's, lie within these limits. We combine RRab's from DR13 along with new CSS
Rab's and MLS RRab's in the range $145\arcdeg < \alpha < 210\arcdeg$ within the Sgr streams region. We transform the
RRab distances to Galactocentric values and plot the sources in Figure \ref{DistG}.  In this plot the distant stream of
RRab's becomes clearer, as does the leading and trailing Sgr streams that are observed between $150\arcdeg < \alpha < 250 \arcdeg$ 
and $10\arcdeg < \alpha < 70\arcdeg$, respectively.

Among the set of distant MLS RRab's, the four most distant objects are found within the $\rm\sim 4\, deg^2$ region
$113.5\arcdeg < \alpha < 115.7\arcdeg$, $22.8\arcdeg < \delta < 24.8\arcdeg$. These RRab's have distances $119.9 < d_{G}
< 129.7$ kpc, their average magnitudes varying by $\sim 0.2$ mags and are consistent with no difference at their
observed brightness. This concentrated group is significantly fainter than any of the other MLS RRab's, suggesting a
possible association between them that varies from the other sources. However, measurements of
radial velocities and metallicities are required before we can rule out association with the other RRab's.

After excluding the group of four very distant RRab's, and four outlier-RRab's near $d_G \sim 70$kpc with $\alpha \sim 115\arcdeg$
we determine the slope of the outer stream by selecting the RRab's with $60 < d_{G} < 115$ kpc, in the range $100\arcdeg
< \alpha < 160\arcdeg$.  A simple linear fit gives $d_G = 177(\pm 5) - 0.68(\pm 0.04) \times \alpha$. The slope of this
line suggests that the distant stream does not meet with the so-called trailing Sgr stream lying in the region 
$0\arcdeg < \alpha < 80\arcdeg$.

\begin{figure}{
\epsscale{0.7}
\plotone{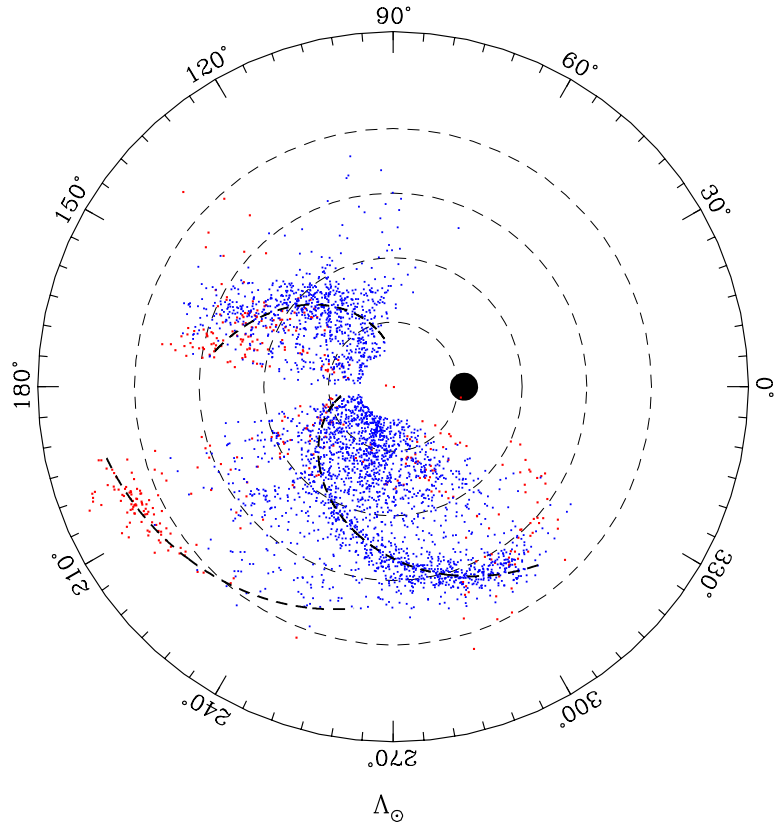}
\caption{\label{Polar}
  A polar plot of the magnitude distribution of MLS (red points) and CSS (blue points) RRab stars in the Majewski et
  al.~(2003) Sgr coordinate system.  The dashed circles present V magnitudes of 17,18,19, and 20.  The three bold arcs
  show the approximate average magnitudes of the three Sgr streams. The large dot shows the location of the Sgr dwarf
  galaxy.
} 
}
\end{figure}

In order to better visualize the direction and slope of the streams, in Figure \ref{Polar}, we present a polar plot of the same
RRab sources in the Sgr plane coordinate system. In contrast to Figure \ref{DistG}, here we plot average static star
magnitudes for the RRab fainter than $V=16.5$ ($d_{H} > 15\rm kpc$). Clearly a more accurate description of path of this
outer stream requires more RRab in the region $\Lambda > 240\arcdeg$, which is not covered by the MLS photometry.  As
noted earlier, many new CSS RRab were found in this region. However, as shown in Figure \ref{MLSHist}, and DR13,
RRab's fainter than $V = 19.5$ ($d_{G} > 60$ kpc) are at the limit of CSS data. Discoveries in CSS data covering this 
region will be likely biased to the brighter, nearer sources. 

\begin{figure}{
\epsscale{1.1}
\plottwo{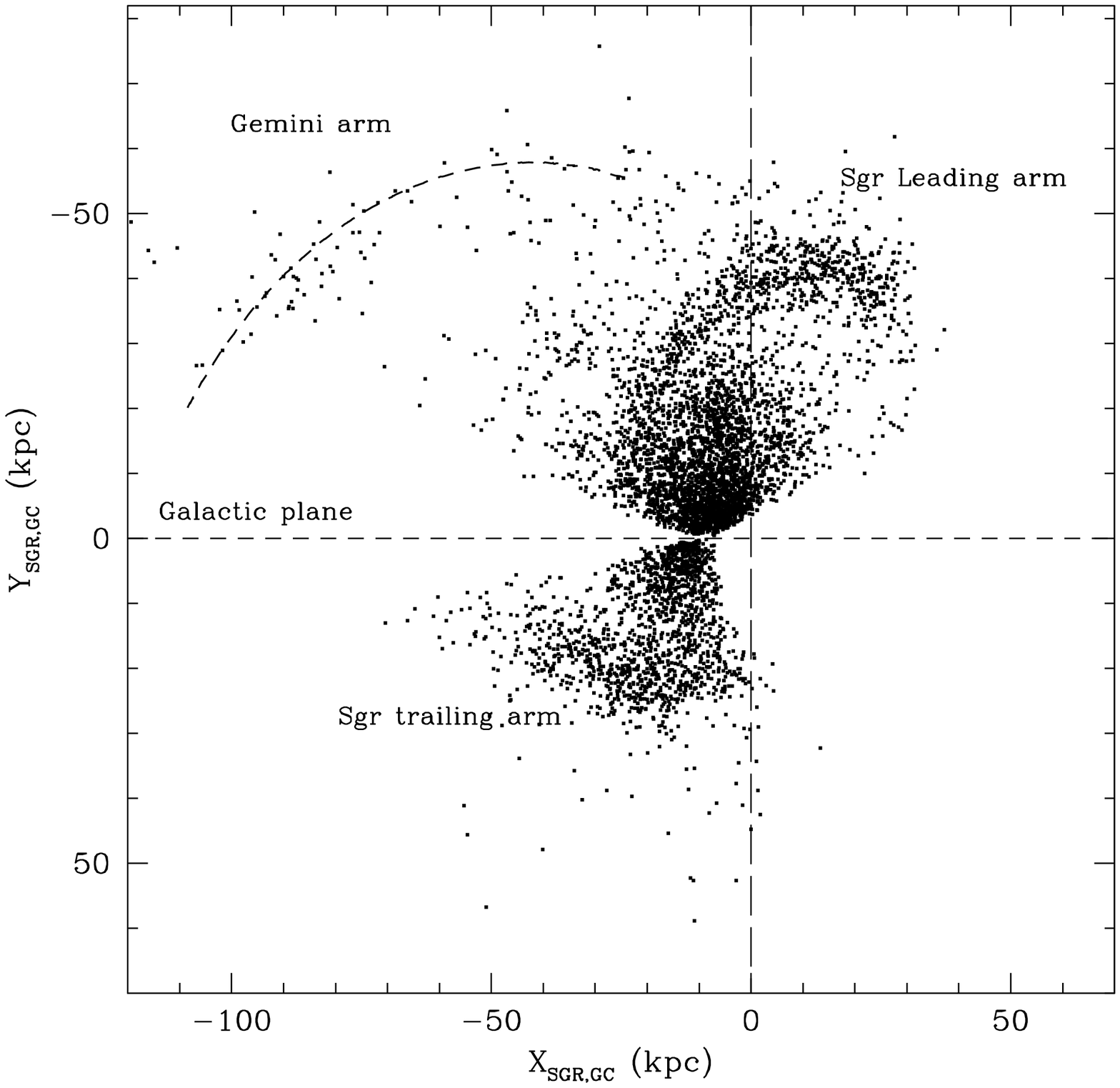}{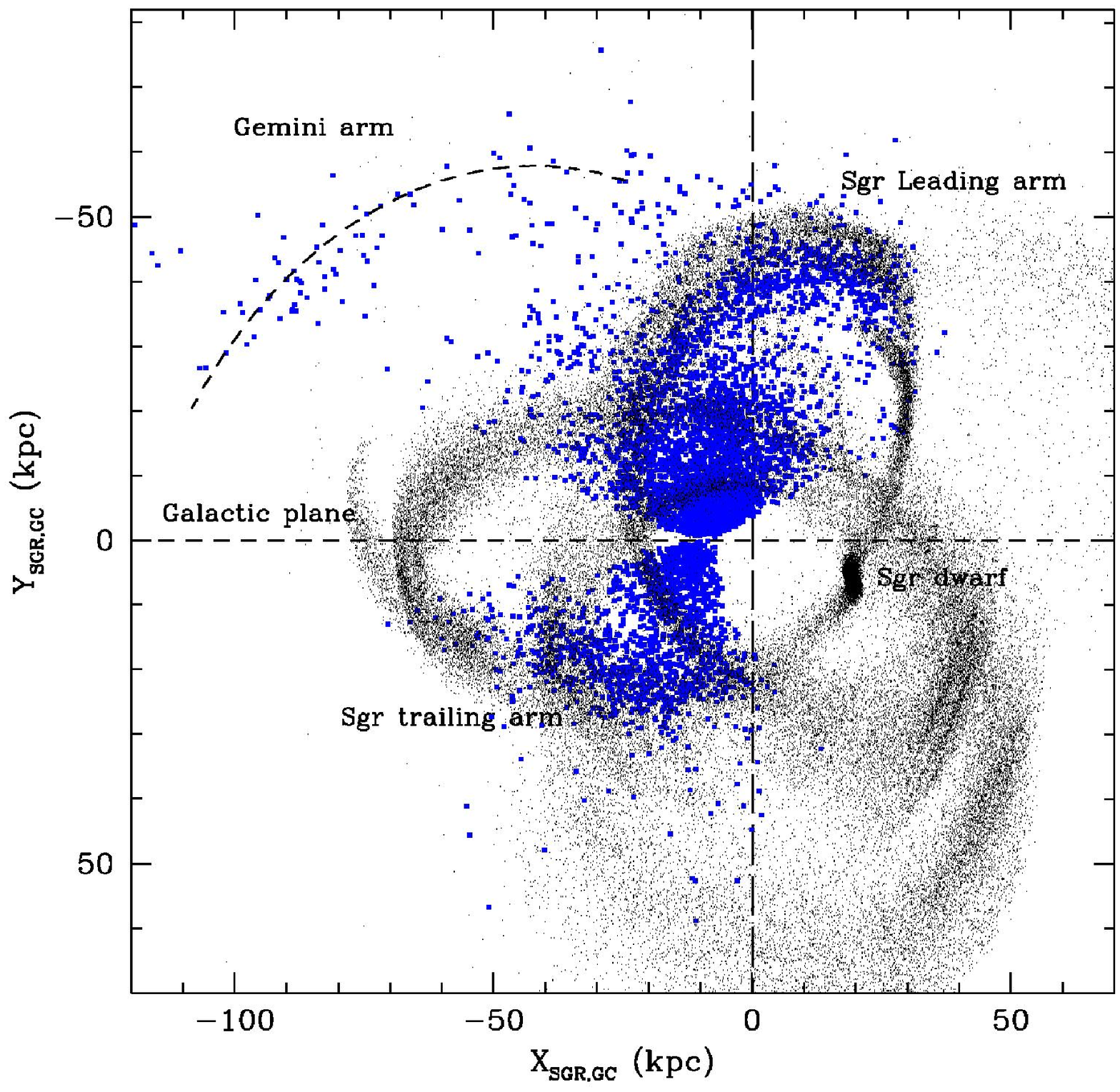}
\caption{\label{SgrXY}
  The distribution of RRL in the Majewski et al.~(2003) Sagittarius coordinate system.  In the left panel we plot the
  distribution of RRab's within $15 \arcdeg$ of the Sagittarius plane. In the right panel we plot the RRab's along with
  the Law and Majewski (2010) Sagittarius streams N-body model.
} 
}
\end{figure}

To further explore the relationship with the Sgr streams, we plot the RRab's in the Sgr X-Y plane system of Majewski et
al.~(2003) in Figure \ref{SgrXY}.  Here we mark the proposed path of the Gemini stream. We also plot the Law \&
Majewski (2010) N-body model of the Sgr stream system. The model is reasonable match to the structure of the inner 
RRab data. The agreement is increased if the source distances are reduced by $\sim 11\%$. 
However, as found by Ruhland et al.~(2011), the Gemini stream is not explained by this model or other
models of the Sgr stream system. Nevertheless, it appears that if the path of Sgr leading debris was
extended, it might better match the location of the sources. Indeed, Law \& Majewski (2010) found that if the
Sgr dwarf had been orbiting 2 Gyr longer, a much longer leading arm would be present. Nevertheless, their simulations did 
not produced sources corresponding to the distances and locations of the BHB candidates and RRL. It is apparent
that the Sgr trailing arm model is a very poor match to the Gemini stream, both in extent and location.

\subsection{Feature Significance}

To determine the significance of the Gemini stream it is necessary to compare the source density with that expected from
halo models.  We adopt the halo density model from Sesar et al.~(2010) for our comparison.  The local density of RRL is
determined using the Nth nearest neighbour method, we find the 8th nearest star to the point where we want to calculate
the density and calculate the area that bounds these stars. We use 8 stars motivated by Ivezic et al.~(2005) where they
found that this number gives the best results for the underlying density on their improved Bayesian method. We also saw
that this number was a good combination of precision and computational efficiency.

\begin{figure}{
\epsscale{0.8}
\plotone{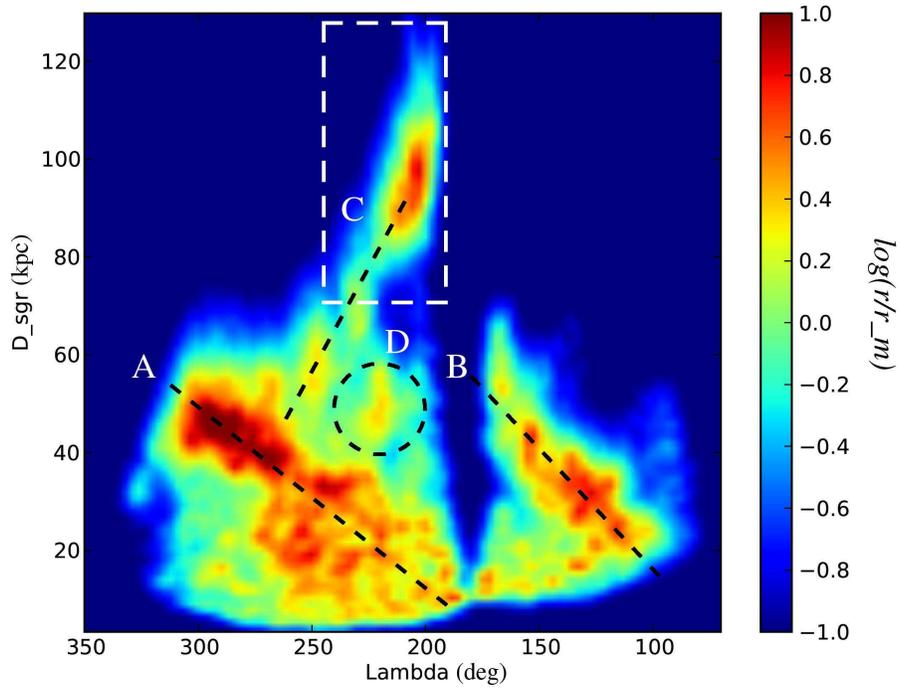}
\caption{\label{SgrGab}
  Halo subtracted density distribution of RRab stars in Majewski et al.~(2003) Sagittarius polar coordinate system.  (A)
  shows the location of RRab's in the Sgr Leading arm.  (B) marks the RRab's in the Sgr Trail arm. (C) shows the
  proposed location of the Gemini stream. (D) marks the location of another possible overdensity of RRab's. The region
  selected for significance testing is outlined by the long-dashed line.
} 
}
\end{figure}

\begin{figure}{
\epsscale{0.8}
\plotone{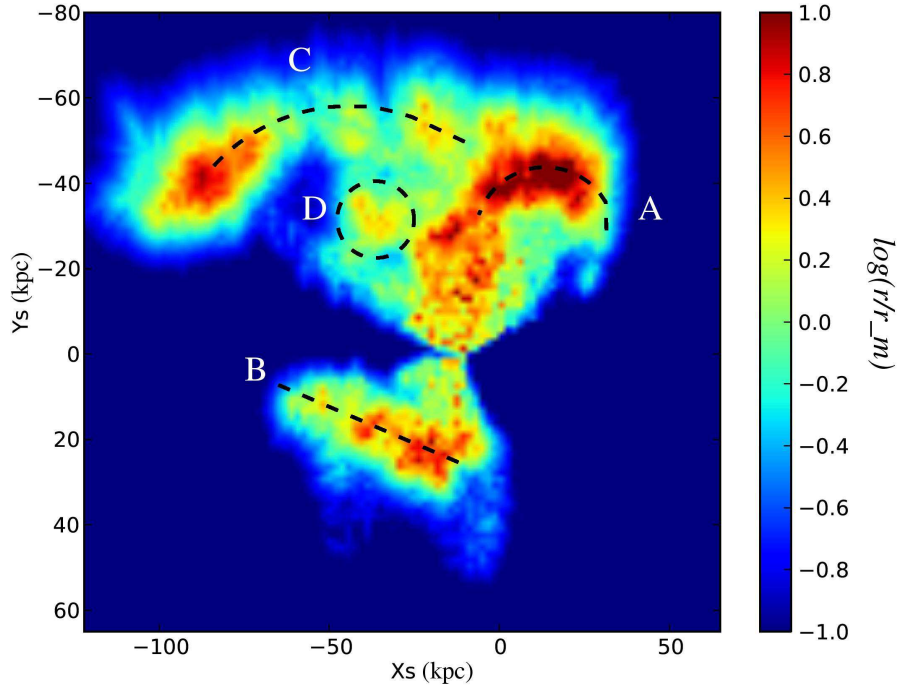}
\caption{\label{SgrGabX}
  Halo subtracted distribution of RRab stars projected to the X-Y plane Majewski et al.~(2003) Sagittarius coordinate
  system.  (A) shows the location of RRab's in the Sgr Leading arm.  (B) marks the RRab's in the Sgr Trail arm. (C)
  shows the proposed location of the Gemini stream. (D) marks the location of another possible overdensity of RRab's.
} 
}
\end{figure}

We compute the observed number density within our grid ($r$), and density from the Sesar et al.~(2010) halo model ($r_{m}$).
To visualize the over-densities on the Sgr plane, we produce a $100 \times 100$ grid on the Mawejeski et al.~(2003) Sgr
coordinates at $Z=0$ (for the X-Y plot) and $B=0$ (for the $\Lambda$-$D_{Sgr}$ plot).  In Figure \ref{SgrGab}, we 
plot the resulting density ratios in the X-Y plane of the Sgr system and in Figure \ref{SgrGabX}, we plot the 
densities in the $\Lambda$-$D_{sgr}$ system. The densities here can be readily contrasted sources in Figure \ref{SgrXY}.

The main density features are the leading and trailing Sgr arms (marked by A and B, respectively).  The Gemini stream
(marked as C), largely has a higher ratio density ratio than the trailing stream system (marked by B).  Another possible
feature is seen near $\Lambda = 220$, $D_{Sgr}=50$ kpc (marked $D$). This has much lower significance than the
other features, yet is lactated where models predict the Sgr trailing arm should cross the Galactic plane and meet
with the Trailing arm $B$. Since the Gemini feature, like much of the Sgr stream, is offset from the Sgr plane, we 
have collapsed the source Z positions to visualize the stream.

Unlike the Gemini stream, the Leading and Trail Sgr arms A and B are all ready well known (Law and Majewski 2010, and
references therein). To quantify the significance of the Gemini stream, we estimated the number of stars expected by the
model at the position of the stream.  For simplicity, we selected the volume that bounds the stream in the Sgr
coordinate system.  That is $190 < \Lambda < 240$, $\rm 1 < B < 17$ and $70 < R_{\rm Sgr} < 130$. Based on the Sesar et
al.~(2010) model density, within this volume we would expect to find 373 RRL assuming 100\% detection efficiency. 
However, only 106 RRL were found.  Assuming pure Poisson uncertainties, this suggests the Gemini area has $\sim
14\sigma$ under-density. As our detection efficiency is close to $70-80$\% based on DR13 this short fall cannot be 
explained pure by missing RRL. The detection of no distant RRL across most of the MLS survey area strongly 
suggests that the halo is much less dense than expected by the model.
This result is in agreement with the results of Watkins et al.~(2009) and Sesar et al.~(2010)
based on halo RRL discovered in SDSS stripe-82. this confirms that the halo density declines 
more rapidly than suggested by the Juric et al.~(2008) halo model.

\begin{figure}{
\epsscale{1.1}
\plottwo{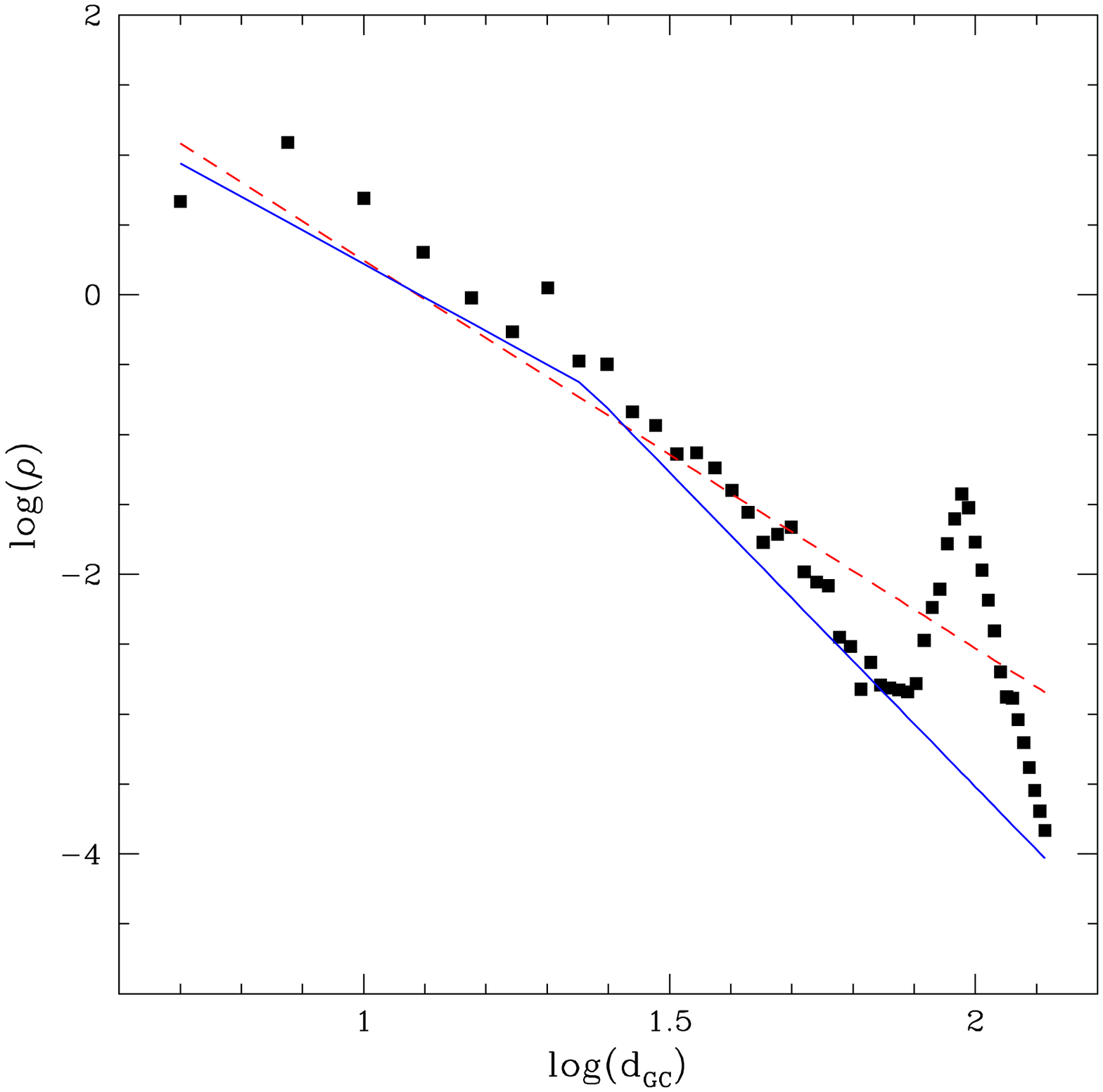}{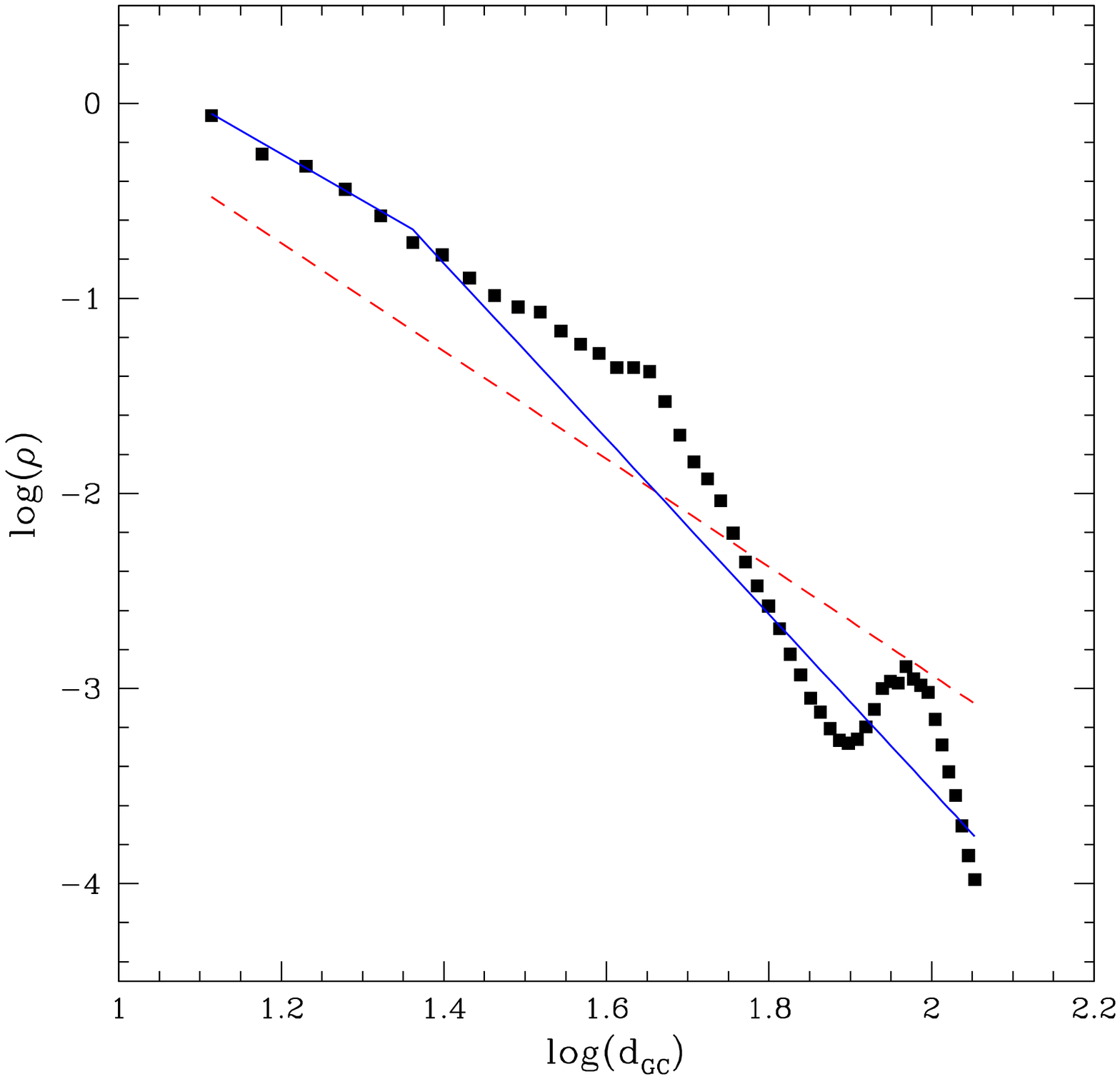}
\caption{\label{SgrGabWat}
  Halo density of RRL.  Here we plot the observed density of RRL compared with the Sesar et al.~(2010) halo model
  (dashed line) and the adjusted Watkins et al.~(2009) model (solid line). In the left panel we plot the values for the
  Sgr along the line $\Lambda=210 \arcdeg$. In the right panel we plot the values averaged over Sgr stream region.
} 
}
\end{figure}

As a second model comparison, we compare the RRL densities with the Watkins et al.~(2009) halo model. Their 
results suggest a break in density occurs $d_{GC} = 23$ kpc. Matching the model to the data we find that the
Watkins et al.~(2009) model requires normalization by a factor of 11 to match our data. This is in 
good agreement with the factor of 10 found by Sesar et al.~(2010).
In Figure \ref{SgrGabWat}, we plot the observed radial density compared with the Sesar et al.~(2010) model 
and the normalized Watkins et al.~(2009) model. The observations are found to be in good agreement 
with the Watkins et al.~(2009) model, although the break in density appears to occur nearer to 50 kpc when
averaged over a large area. However, this is because the densities are enhanced by RRL in the Sgr 
leading and trailing arms. Based on the Watkins et al.~(2009) model we expect to find 50 RRL in the 
region selected above. The number of Gemini stream RRL we find is thus $7.9 \sigma$ larger than
expected from this model.

\subsection{RR Lyrae Populations}

To investigate differences between the Gemini RRab sources and the overall population, one can infer the
Oosterhoff type based on the RRab period-amplitude relationship (Smith et al. 2011, and references therein). However,
since average colours were used to transform the MLS light curves, we need to account for the effect of colour variation
of the RRab light curves.  In DR13 we found that RRab amplitudes were systematically reduced by 0.15
due to pulsational colour variations within the broad bandpass of CSS images compared to V-band.
We also found that the CSS Oosterhoff type-I (OoI) RRab's exhibit a well-defined amplitude limit that is 
0.1 mags higher than the Zorotovic et al.~(2010) period-amplitude relationship.

\begin{figure}{
\epsscale{1.1}
\plottwo{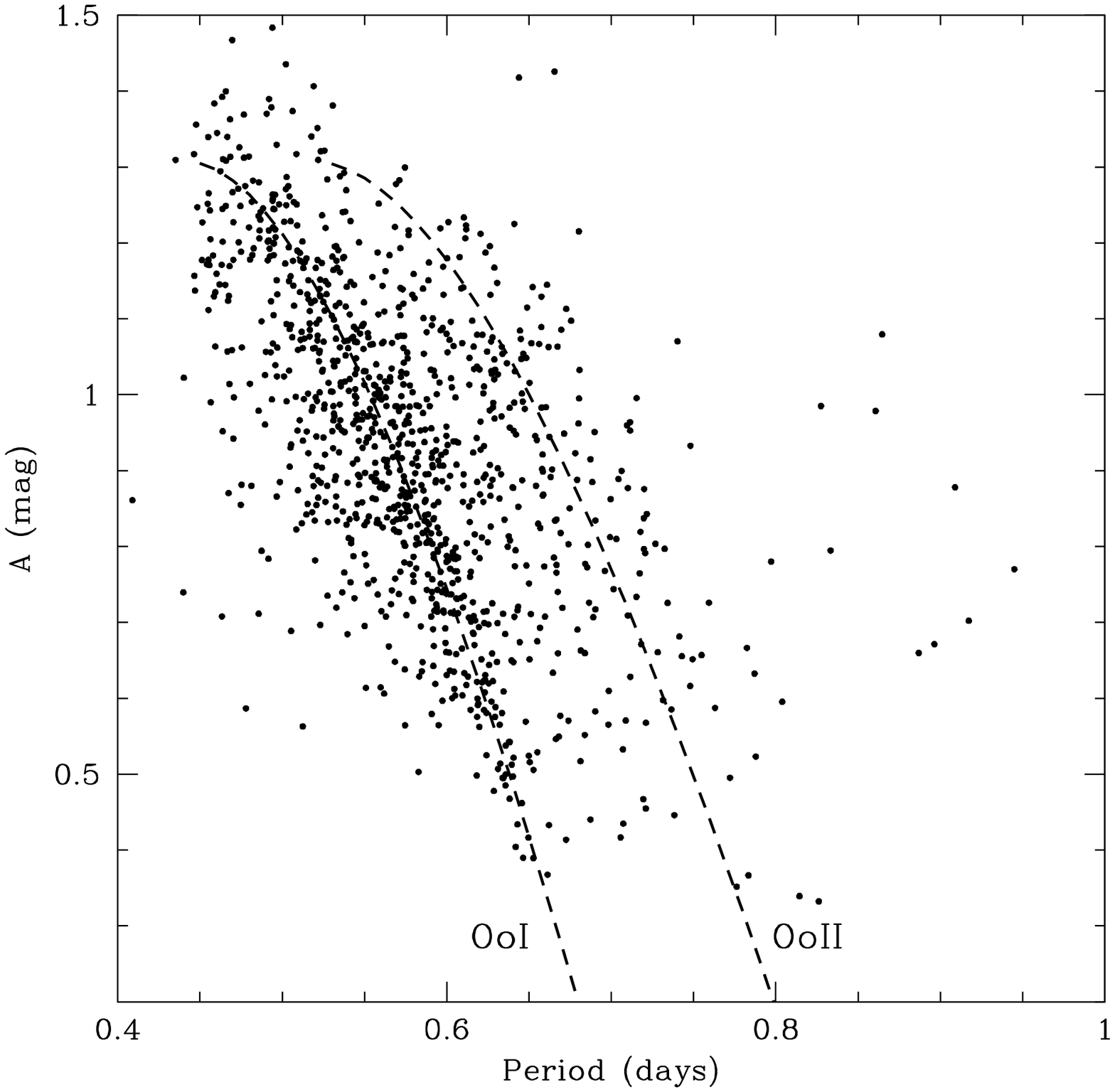}{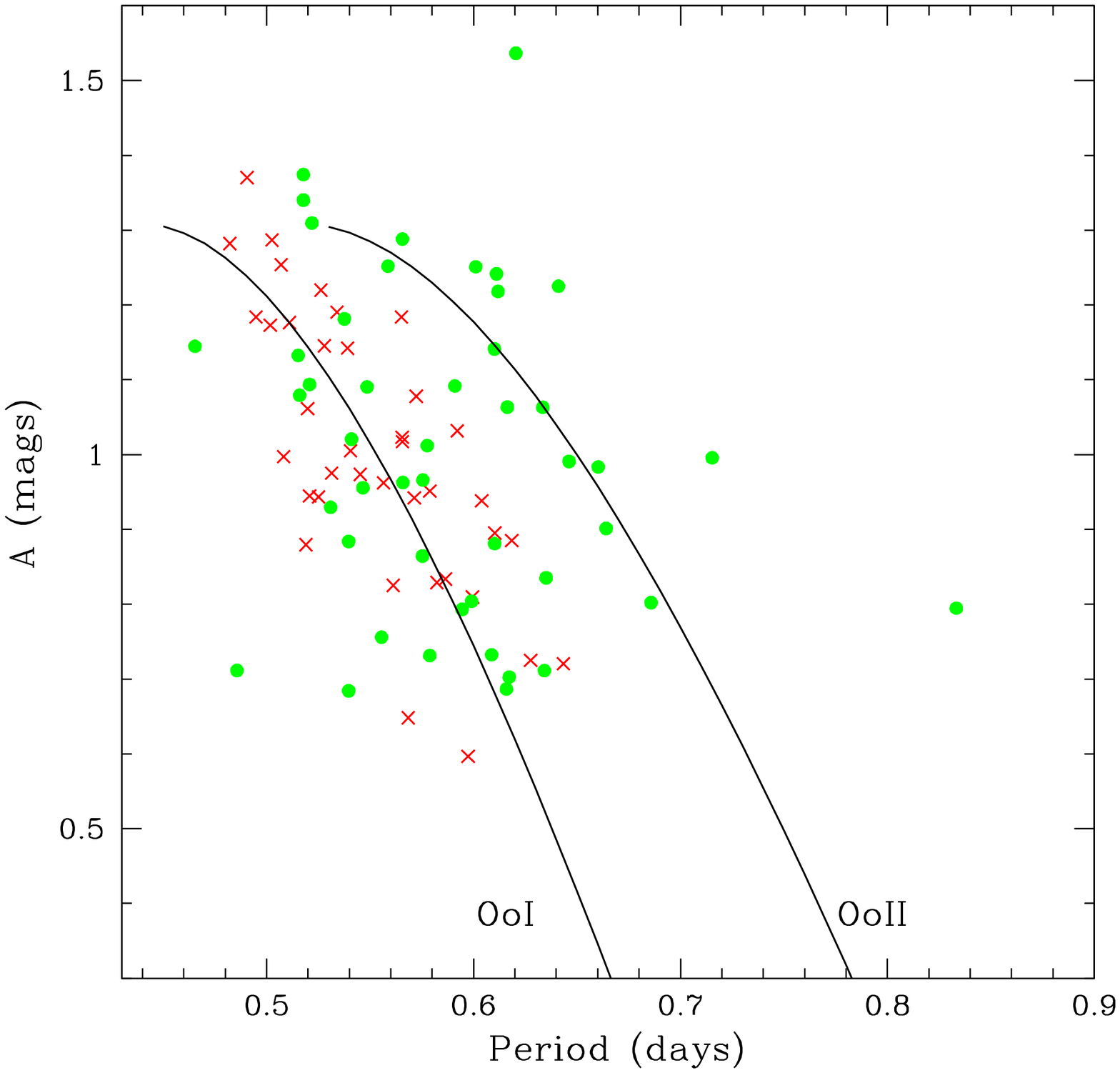}
\caption{\label{MLSOtype} 
  The period-amplitude distribution of RRab stars.  In the left plot we present the period and amplitudes for all the
  MLS RRab's.  In the right plot, the MLS RRab's with $70 < d_G < 95$ kpc are plotted with green dots, and those with
   $d_G > 95$ kpc are red crosses.  The solid-lines mark the Zorotovic et al.~(2010) period-amplitude relationships for 
  OoI and OoII and the dished-lines outline a region where the RRab Oosterhoff types are mixture of OoI and OoII. 
  shown. In both plots the RRab V-band amplitudes plotted have been increased by 0.15 mags to account for colour 
  variations in the RRab's between maximum and minimum light.
}
}
\end{figure}

To separate OoI RRab candidates from Oosterhoff type-II's (OoII's), we correct the MLS amplitudes to values by adding
0.15 mags. Next we define a set of RRab's that are between 0.1 and 0.25 mags above than the average OoI period-amplitude
relationship.  These objects are a mixture of Oosterhoff types (although some may be Oosterhoff-intermediate sources)
that we remove from consideration. We then investigated the variation in MLS RRab's Oosterhoff types with distance.
In Figure \ref{MLSOtype} we plot the period-amplitude distribution of the MLS RRab's.

By dividing these sources into a nearer sample consisting of RRab with $70 < d_G < 95 \, {\rm kpc}$ and the more distant
RRab $d_G > 95 \, {\rm kpc}$ we see a division in types.  In Figure \ref{MLSOtype}, we also plot the period-amplitude
distribution for the Gemini RRab's. We also over-plot the Zorotovic et al.~(2010) OoI line and an OoII line that is offset
by $+0.07$ in $\log(P)$ to match the CSS observations.  The nearer RRab exhibit a mixture of Oosterhoff types when
compared to the Zorotovic et al.~(2010) period-amplitude relationships.  This is significant evidence that the RRab's in
the range $70 < d_G < 95 \, {\rm kpc}$ do not come from a single population.  In contrast, the distant MLS
RRab's all lie near the OoI line, suggesting that they come from a single population that is more metal-rich.

\begin{figure}{
\epsscale{0.6}
\plotone{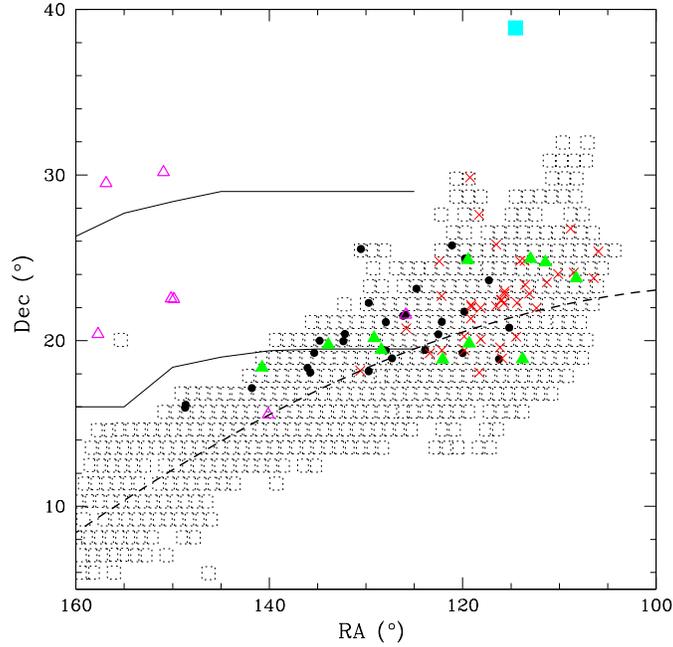}
\caption{\label{MLSHBd}  
  Locations of distant halo RRab's.  Here we present the locations of RRab's relative to the MLS survey fields marked by
  the open blue boxes.  The RRab with $70 < d_G < 95\,\rm kpc$ and periods and amplitudes most consistent with
  Oosterhoff-I (i.e., Sagittarius dSph-like) sources are marked with large dots. The MLS RRab in this distance range
  that are most consistent with an Oosterhoff-II type (i.e., NGC~2419-like) classification are given by filled green
  triangles. The CSS RRab's with $70 < d_G < 95\,\rm kpc$, but uncertain Oosterhoff types are marked with open magenta
  triangles.  The MLS RRab's with $d_G > 85 kpc$ are given by red crosses. Here the short-dashed line shows the location
  of the ecliptic plane.  The large cyan square shows the location of NGC 2419.  The solid lines shows the rough
  locations of two Sgr streams given by Belokurov et al.~(2006).  
} 
}
\end{figure}

In Figure \ref{MLSHBd}, we show the locations of Gemini stream RRab's after removing the MLS RRab's with ambiguous 
Oosterhoff type. The CSS RRab's with $70 < d_G < 95 \, {\rm kpc}$ have also been included, but are not separated 
by Oosterhoff type since their amplitudes and periods are less certain than the MLS RRab's.  This figure shows that 
the Gemini RRab's are spread across the width of Sgr stream system. It also shows that the OoII RRab's selected 
among the nearer RRL set ($70 < d_G < 95 \, {\rm kpc}$) are distributed across the region. The nearer OoI group ends 
around $\alpha \sim 125\arcdeg$ where the stream divides between the nearer and further groups. Analysis of
this figure suggests that there may be two or more overlapping populations: one that follows a steep distant 
gradient and is predominantly OoI, and another that has a shallower gradient and is an OoII population. 
With respect to this figure, we once again note that the MLS RRL are naturally concentrated toward the 
ecliptic because of sampling. So it is not possible to make inferences about changes in density 
across the Sgr stream.

\subsection{Comparison with SDSS data}

While the SDSS photometry has little of the repeated photometry required to unambiguously identify RRab's, 
it is deeper than both CSS and MLS data (reaching HB stars to $g \sim 22$). Additionally, SDSS data does
cover most of the Gemini stream observed in the MLS data as well as the main region where MLS data 
does not overlap the Sgr stream. Therefore, it is possible to use this data to bridge the gap in between
the depth of CSS data and coverage of MLS photometry. 

Unlike previous authors whom searched for BHB stars covering the Sgr tidal streams, here we sought to select both BHB
and potential RR Lyrae stars.  Based on the colours of the distant MLS RRab's and prior BHB work, we investigated the 
observed colours of RRL in SDSS photometry and selected SDSS DR8 stars within
$0.95 < (u - g)_0 < 1.5$, $-0.2 < (r - i)_0 < 0.2$, $-0.35 < (g - r)_0 < 0.22$, SDSS object type=6 (star), and $17 < g_0 <
22$.  Based on DR13 we know that the bulk of RRab's are located near $(g - r)_0 = 0.25$, $(r - i)_0 = 0.1$ and $(i-z)_0
= 0.05$.  However, RRab's with these colours are outnumbered by main sequence turnoff stars (MSTO) by a factor of $>
100$ (Koposov et al.~2012). Using maps of SDSS source density, we sub-selected stars in the range 
$2.7\times (r-i)_0 + 0.25 > (g - i)_0 > 2.7\times (r-i)_0 - 0.1$ for $(g -i)_0 < 0.05$.  This selection retains 
most of the BHB candidates of Ruhland et al.(2011) as well as a 22\% of the distant MLS RRab sample and many BS stars.

\begin{figure}{
\epsscale{0.65}
\plotone{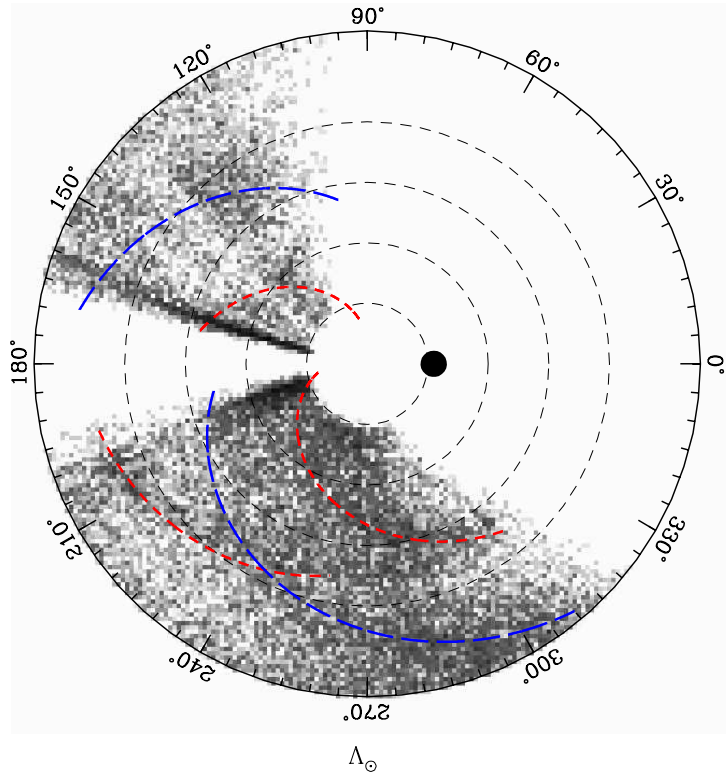}
\caption{\label{Hess}
  The spatial point-density distribution of SDSS HB candidates near the plane of the Sgr streams ($-11\arcdeg < B <
  11\arcdeg$) with magnitudes $17 < V < 21.5$.  The dashed circles appear at magnitudes $V=17$, 18, 19, and 20.  The
  locations of RRab streams from Figure \ref{Polar} are presented with thick red short-dashed lines while the expected
  locations of BS streams which mirror the HB streams are shown with thick blue long-dashed lines.
}
}
\end{figure}

This colour selection reduces the initial number of SDSS point sources in our selection from 1.65 million to 81552.  By
selecting stars in the Sgr stream region, with $-11\arcdeg < \Lambda < 11\arcdeg$, the number reduces to 23507 HB
candidates. To match the MLS photometry we transform the SDSS photometry to $V$ magnitudes using Ivezic et al.~(2007). 
As found with Newberg et al.~(2003, 2007) and Ruhland et al.~(2011) we found the spatial distribution of SDSS HB
candidates suffers from a significant crowding and background that is best viewed in source density. In Figure \ref{Hess}, 
we provide the HB candidates in the form of the Hess (point-density) diagram. 
This figure shows evidence that the density of HB candidates closely follows the fit to the distant Gemini stream 
of RRab's, in agreement with the distribution of BHB candidates from Ruhland et al.~(2011) and others.
Here the scaling has been set to match that in Figure \ref{Polar} where the main streams are visible.
In addition to the RRL, the paths expected for BS stars that pass the colour cuts and form a shadow that
is $\sim 2$ mag fainter than the HB stars is also shown.

\begin{figure}{
\epsscale{1.0}
\plotone{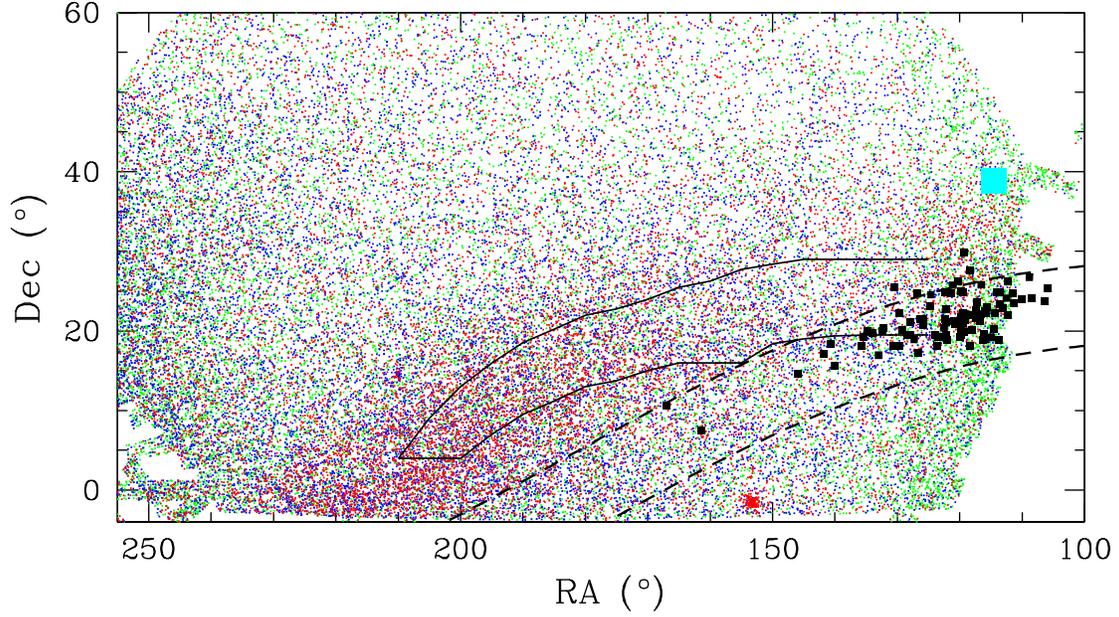}
\caption{\label{MLSHB} 
  The distant MLS RRL and SDSS HB candidates overlapping the Sgr tidal streams.  The green points show SDSS stars with
  $17 < V < 18.5$, the blue points those with $18.5 < V < 19.8$ and the red points those with $19.8 < V < 20.7$.  The
  black squares show the locations of RRL with $d_G > 80\rm\, kpc$, and the large cyan box shows the location of NGC
  2419.  The dashed-lines connect points with $b= -10$ and $b = 10$.  These are the approximate limits of the MLS
  survey. The solid lines show the locations of two Sgr streams given by Belokurov et al.~(2006).
}
}
\end{figure}

To obtain another view of MLS RRL in the Gemini stream in relation to the SDSS HB candidates, we divided the SDSS
sources into three groups. These were bright sources, with $17 < V < 18.5$, intermediate-brightness sources, with $18.5 <
V < 19.8$, and faint sources, with $19.8 < V < 20.7$.  In Figure \ref{MLSHB}, we present the locations of these sources.
The bright sources were selected to show the nearby HB stars as well as the BS stars in the Monoceros stream (near the 
Galactic anti-center at $\alpha=110-120\arcdeg$ limit of the SDSS coverage). The intermediate-brightness sources were 
selected to be indicative of the HB stars in the leading arm of the Sgr stream. The faintest sources were selected to 
include the HB candidates in the Gemini stream. However, this also include BS stars that mirror the distribution of HB 
stars along the Sgr leading arm. The figure clearly shows the overdensity associated with the Sgr streams system. The 
separation into two streams is not clear here since there are far fewer BHB stars than the MSTO stars used by 
Belokurov et al.~(2006).

\subsection{SDSS Spectra of RRL Sources}

We matched the entire MLS RRL catalog with the SDSS DR8 spectroscopic catalog and found 89 matches.  A much
larger sample of RRL with spectra is given in DR13. However, here our main purpose was analysis of the distant halo RRL.  
We found that 16 of these matches are sources beyond 70 kpc. As noted earlier, many of the faint RRL candidates were
found based on their SDSS spectra. The 16 RRL with SDSS spectra include 12 RRab's and four RRc's.

To separate the radial velocities from the velocities due to pulsation, we use the SDSS observation times and the
Fourier fits to derive the phase at which the RRL spectrum was observed.  As noted in DR13, for SDSS spectra, radial
velocities are determined by averaging both Balmer and metallic lines (mainly Ca lines). We follow DR13 by applying 
Sesar et al.~(2012b) velocities corrections for pulsation based corrections derived from both hydrogen and metallic
lines. The average velocity correction for the 12 RRab is $\rm 13~km/s$, in good agreement with the uncertainty 
derived from 905 RRab spectra in DR13 ($\sigma= 14.3$ km/s).  We adopt this level of uncertainty for all the SDSS 
RRab spectra.

For the four RRc stars which pulsate in the first overtone mode the velocity of the pulsation is much smaller than for the
RRab's. Based on the RRc's observed by Liu \& Janes (1989) and Jones et al.~(1988), the amplitude is expected to be
approximately $\rm 20~km/s$. To account for this factor, for these RRL we increase the observed SDSS radial velocity
uncertainties by an additional $\rm 10~km/s$.  Following Law \& Majewski (2010) we transform the radial velocities to the
Galactic standard of rest assuming a Solar peculiar motion of (U, V, W) = (9, 12 + 220, 7) km/s in the Galactic
Cartesian coordinate system.

\begin{figure}{
\epsscale{1.0}
\plotone{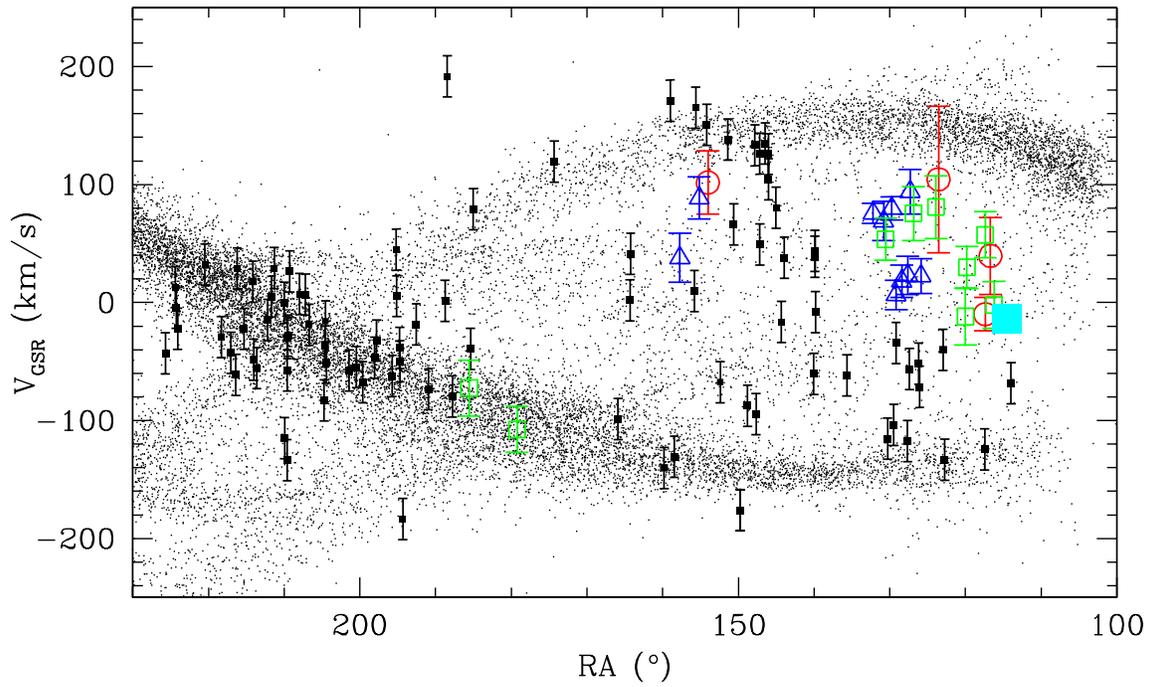}
\caption{\label{MLSVel}   
  Radial velocities for outer halo RR Lyrae. The blue triangles are values for Sesar et al.~(2012a) Cancer Group A" and
  "Cancer Group B" RR Lyrae,.  Green squares and red circles are MLS RRab and RRc measurements, respectively.  The dots
  present points from the Law \& Majewski~(2010) Sgr N-body model. The large filled cyan box presents the radial
  velocity and right ascension of NGC 2419.  The black squares are CSS RRab's with distances $d_G > 40 kpc$.
}
}
\end{figure}

In Figure \ref{MLSVel}, we plot the radial velocities for the outer halo MLS RRab's with spectra, the CSS RRab's with
$d_G > 40$ kpc, and the 10 PTF RRab's found by Sesar et al.~(2012a). We also plot the velocities predicted by Law \&
Majewski (2010) N-body simulations. The figure shows that two of the RRab's observed near $\alpha =190\arcdeg$ appear to
be associated with the leading arm of the Sgr stream.
However, for the other Gemini stream RRab's, the velocities, like the distances, are not explained by the Law \&
Majewski (2010) N-body simulations. In this regard, Sesar et al.~(2012a) noted that the RRab velocity measurements they
found suggested the RRab's belonged to two distinct groups. Additionally, sources associated with our group-D 
of Figure \ref{SgrXY} (occurring near $\alpha=150\arcdeg$) have radial velocities that appear to vary rapidly with 
Right Ascension. These velocities are also inconsistent with the Law \& Majewski (2010) model, suggesting that they 
do not belong to the Sgr leading arm. The association between these feature-D RRab's and the Gemini stream RRab's
is unclear, although values are similar.

\subsection{Links to NGC~2419}

As noted by Newberg et al.~(2003), the outer halo BHB candidates they discovered reside near the unusual
Globular cluster (GC) NGC~2419.  This system is located at $\alpha=114.53\arcdeg$, $\delta=38.88\arcdeg$ and distance
$d_h=82.6$ kpc (Harris 1996, 2010 edition). The corresponding coordinates in the Majewski et al.~(2003) Sgr 
coordinate system are $\Lambda =201.7\arcdeg$, $B=-8.5\arcdeg$, making it well within the limits of the Sgr stream 
system (Koposov et al.~2012). The Galactocentric radial velocity is given by Newberg et al.~(2003) as $\rm -14 km/s$, and  
Baumgardt et al.~(2009) find an internal velocity dispersion of $\rm 4 km/s$.

NGC~2419 is noted as being one of the most metal-poor GCs ($\rm [Fe/H] \sim -2.1$; Mucciarelli et al.~2012).  The
cluster is notably old, with age 12.3 Gyr according to Forbes \& Bridges (2010).  The system has the highest luminosity
($M_V \sim -9.6$ mag) of any GC with a galactocentric distance $R > 15$ kpc, apart from the likely Sgr dwarf-associated 
GC, M54 (Cohen et al.~2010). The half-light radius of this cluster is 19 pc making it significantly more extended
in the $\log(R_h)$ versus $M_{V}$ plane than other GCs with $R > 15$ kpc (Mackey \& van den Berg 2005).
Indeed, the exceptional nature of NGC~2419 relative to outer halo GCs led van den Berg \& Mackey~(2004) to suggest that
the object is the stripped core of a former dwarf spheroidal galaxy (dSph).  Based on abundance studies Cohen et
al.~(2010) and Cohen \& Kirby (2012) also found that the NGC~2419 appears like no other globular cluster, but rather 
the core of an accreted dwarf galaxy.

From Figures \ref{DistG} \& \ref{MLSVel} we see that the location and radial velocity of NGC~2419 are a relatively good
match for the Gemini tidal stream. The average velocities and metallicities of the Sesar et al.~(2012a) Cancer
group B ($\rm {\langle v \rangle}_{gsr} =16.3\pm7.1~km/s$) and $\rm [Fe/H] = -2.1 \pm 0.4~dex$) and the metallicity 
of the Cancer group B is in reasonable agreement with the values expected for a stream from NGC~2419. However,
the velocities and metallicities do not provide a strong enough association to link these sources.

Another important point to consider in the possible association between the Gemini RRL and NGC~2419 is the proximity 
of the sources. Some of the RRL have $\delta < 20\arcdeg$ near the Right Ascension of the NGC~2419
($\alpha=114.53\arcdeg$).  Thus, if one was to assume that the Gemini tidal stream proceeds in the direction 
of the Sgr stream (as suggested by SDSS HB candidates), the stream stars would have to be dispersed across the 
entire $\sim 20\arcdeg$ of the Sgr system between the MLS RRab's and NGC~2419. Sources within this gap are not 
covered by MLS observations because of the coverage limits of the survey. It is thus remain unclear whether 
there is any link between the Gemini stream and NGC~2419. 
Also, as noted earlier, our data shows that the most distant Gemini RRab's are OoI type stars while 
NGC~2419 is well known to be OoII type. Indeed, Mucciarelli et al.~(2012) found that NGC~2419 exhibits very 
little spread in $\rm [Fe/H]$ ($\sigma = 0.11$ dex) suggesting it could not be linked with metal-rich RRL 
in the Gemini structure.

\section{Discussion and Conclusions}

We have performed a periodicity analysis of 3.1 million variable star candidates selected from photometry taken by the
MLS survey and uncovered 1,207 RRL (of which 538 are new). Comparison of the periods for the $\sim 600$ previous known
RRL shows that the sources are accurately measured. We have also discovered 2040 new RRab stars in a re-analysis
of CSS photometry.

Using Catalina Surveys photometry we have determined the nature of 90\% of the SEKBO (Keller et al. 2006) RRL candidates
and find that 60\% are likely to be RRL.  Our analysis of the SEKBO RRL candidates revealed, selection of a pure set of
RRab's with accurate average magnitudes requires many observations.  The importance of repeated observations for
characterizing variable star types was also recently, demonstrated by Sesar et al.~(2010), who used RRab
light curves to study over-densities that they had earlier been attributed to RRL in Sesar et al.~(2007).  They discovered
that a number of the over densities attributed to RRL, based on photometric selection coupled with a small number of
observations, were in fact due to intrinsically fainter $\delta$ Scuti stars as well as non-variable sources. 
Overall they found that only 70\% of their initial candidates were RRL stars.  Similarly, Ivezic et al.~(2005) found
that although it was possible to completely colour-select a small fraction (6\%) of RRL based on SDSS photometry,
if one wanted to select 60\%, a 72\% non-RRL contamination rate would result. Similar levels of contamination make
it equally difficult to trace halo structures using photometrically selected BHB stars.

In our analysis we also found a significant group of RRL with average $V$ magnitudes $\sim 20.5$.  By combining these 
sources with CSS RRab's, and photometrically-selected HB candidates, we find strong evidence for a tidally disrupted 
stellar stream that crossing $> 60\arcdeg$ of the sky at Galactocentric distances from 70 to 110 kpc. This result 
confirms the existence of a stream first noted by Newberg et al.~(2003). Comparison with halo density models
shows that the feature is significant and that the halo density declines rapidly beyond 30-50 kpc, as previously noted
by Watkins et al.~(2009) and Sesar et al.~(2010). However, since these results, like those of 
Watkins et al.~(2009) and Sesar et al.~(2010), are based on a thin slice through the halo, caution 
has to be taken when interpreting the extent of density regions outside the observed fields.

Although the Gemini RRL overlap with the Sgr stream system, we find that the large distances are inconsistent 
with existing Sgr models. This result is in agreement with Ruhland et al.~(2011) and a recent sample of RRL 
discovered by Sesar et al.~(2012a).  Furthermore, we find that the radial velocities of the RRL are inconsistent 
with simulations of Sgr stream.  However, we note that models of the Sgr streams system as a whole remains 
poorly constrained by observations. 

We have investigated the possible relationship between the Gemini tidal stream and NGC~2419 as first proposed by Newberg
et al.~(2003). The most recent analyses of NGC~2419 shows significant evidence for it being the nuclear remnant of a
disrupted dwarf galaxy (van den Berg \& Mackey 2004; Mackey \& van den Berg 2005; Cohen et al.~2010, 2011; Forbes \&
Bridges 2010; Cohen \& Kirby 2012).  Although, we find that the distances of many of the RRLs and HB candidates are
consistent with NGC~2419, the available velocities and location of the Gemini stream are insufficient agreement to link
the two structures.  Furthermore, the most distant of the RRab's discovered appear to be metal rich sources and would
thus be inconsistent with stars observed in NGC~2419.  Nevertheless, given the location NGC~2419 within the halo, and
with 10 degrees of the Sgr streams plane, it seems possible that a stream associated with NGC~2419 could join it to the
Sgr system.  This may in part account for the significant diversity in metallicity observed for varying Sgr stellar
streams (Law \& Majewski~2010).  Furthermore, as the photometric-selected HB candidates and RRab's cover the Sgr stream
system and exhibit a distance gradient, it is possible that there is a second galaxy remnant associated with the Sgr
stream. Such a source might explain the origin of the two intersecting streams of the Sgr system that has now
been well delineated with MSTO stars by Belokurov et al.~(2006) and Kosopov et al.~(2012). However, even with 40\%
uncertainties in the distances to MSTO stars (Newby et al.~2011), the Gemini stream RRab's are twice as distant as
expected for MSTO stream stars (Koposov et al. 2012). Alternately, the Gemini stream may originate from the remnant 
of another disrupted dwarf galaxy that lies beyond the Gemini stream stars and is yet to be discovered. The 
Gemini stream leads into the galactic plane beyond 100kpc. A highly extincted system in the Galactic plane 
would be very difficult to detect.

Future photometric and spectroscopic observations of the HB stars within $10\arcdeg$ of NGC~2419 could confirm whether
there truly is a tidal stream of RRL associated with NGC~2419. For example, the seemingly unique Mg and K abundance
patterns in NGC~2419 found by Mucciarelli et al.~(2012) would chemically tag stars originating from this source, even in
the presence of overlapping tidal streams. If the Gemini tidal stream does follow the path expected from the SDSS HB
candidates, additional deep photometric observations undertaken by projects such as LSST (Abell et al.~2009) should
reveal numerous additional RRL along this tidal stream. Moreover, if these stars are associated with NGC~2419, we
predict that the RRL will mostly be type Oosterhoff II.

\acknowledgements 
CRTS and CSDR1 are supported by the U.S.~National Science Foundation under grants AST-0909182 and CNS-0540369.
The CSS survey is funded by the National Aeronautics and Space Administration under Grant No. NNG05GF22G issued through the
Science Mission Directorate Near-Earth Objects Observations Program. J. L. P. acknowledges support from NASA through
Hubble Fellowship Grant HF-51261.01-A awarded by the STScI, which is operated by AURA, Inc.  for NASA, under contract
NAS 5-26555.  
Support for M.C. and G.T. is provided by the Ministry for the Economy, Development, and Tourism's Programa Inicativa
Cient\'{i}fica Milenio through grant P07-021-F, awarded to The Milky Way Millennium Nucleus; by Proyecto Basal
PFB-06/2007; by Proyecto FONDECYT Regular \#1110326; and by Proyecto Anillo ACT-86.
SDSS-III is managed by the Astrophysical Research Consortium for the Participating Institutions of the SDSS-III
Collaboration Funding for SDSS-III has been provided by the Alfred P. Sloan Foundation, the Participating Institutions,
the National Science Foundation, and the U.S. Department of Energy Office of Science. The SDSS-III web site is
http://www.sdss3.org/.

\newpage
\onecolumngrid

\begin{figure}[ht]{
\hspace*{-2.5cm}\vspace*{0.5cm}\epsscale{1.45}
\plotone{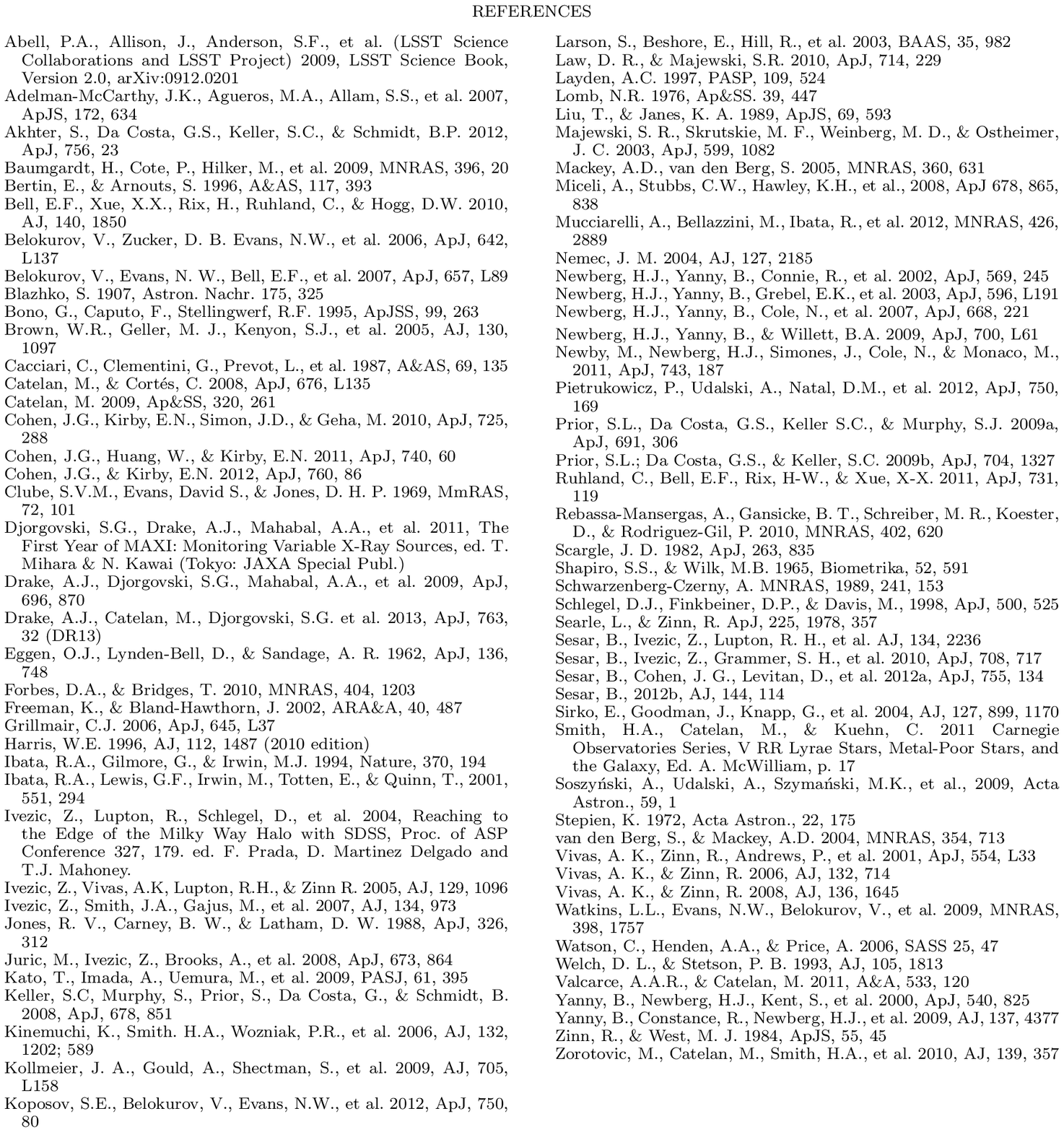}
}
\end{figure}

\newpage

\begin{figure}{
\epsscale{1.2}
\plotone{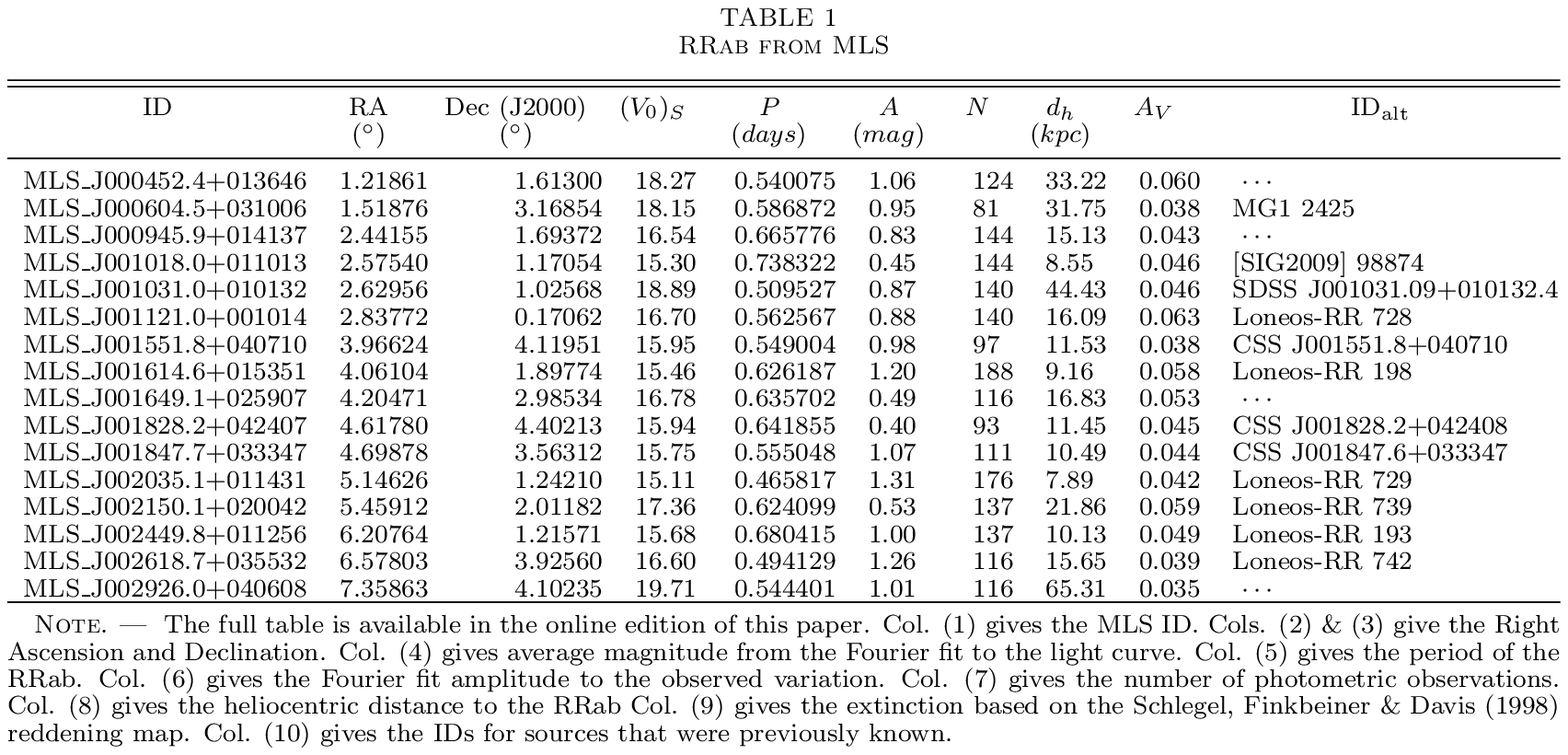}
}
\end{figure}

\begin{figure}{
\epsscale{1.1}
\plotone{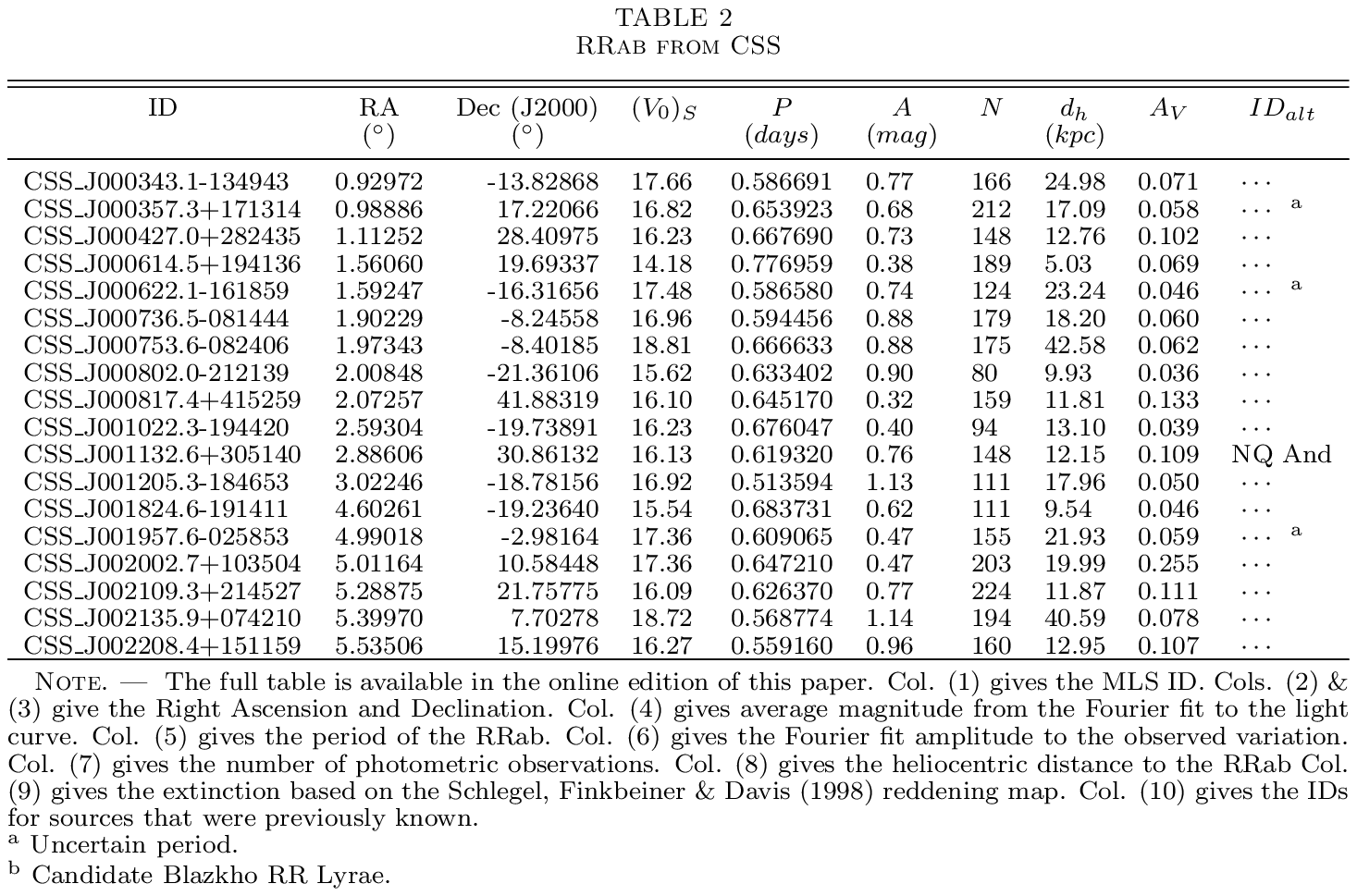}
}
\end{figure}

\begin{figure}{
\epsscale{1.2}
\plottwo{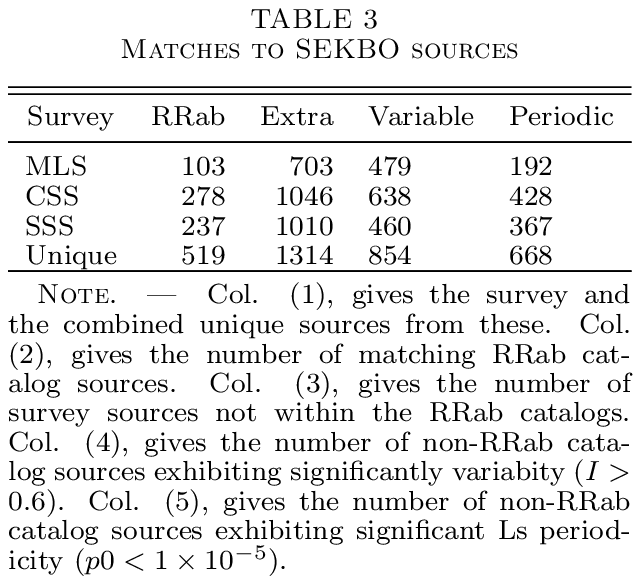}{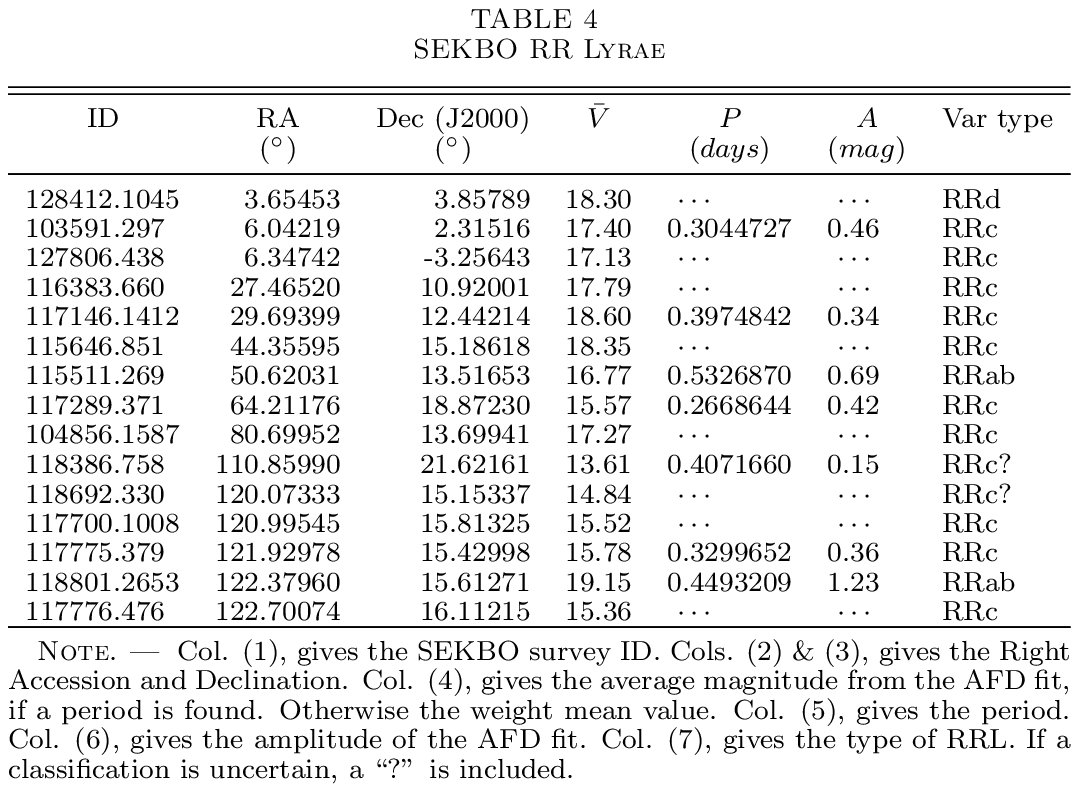}
}
\end{figure}


\clearpage

\clearpage

\clearpage

\end{document}